\documentclass[journal,a4paper]{IEEEtran}
\usepackage{multirow}
\usepackage{graphicx}
\usepackage[table,xcdraw]{xcolor}
\usepackage{balance}
\usepackage{cite}
 \IEEEoverridecommandlockouts
\usepackage{cite}
\usepackage{amsmath,amssymb,amsfonts}
\usepackage{algorithmic}
\usepackage{array}
\usepackage{nccmath}
\usepackage{tabularx}
\usepackage[font=footnotesize]{caption}
\usepackage{caption}
\usepackage{subcaption}
\usepackage{float}
\usepackage{amsmath}
\usepackage{upgreek}

\usepackage{pgfplots}
\usepackage{tikz}
\pgfplotsset{compat=newest}
\usepackage{xcolor}
\usepackage{lipsum}
\usepackage{graphicx}
\usepackage{tikz}
\usepackage{pgfplots}
\usepackage{xcolor}
\usepackage{color, colortbl}
\usepackage{bbm}
\usepackage{booktabs}
\usepackage{amsmath, amsfonts, amssymb, mathtools}

\usepackage{Colori} 
\usepackage{makecell}

\usepackage[acronym]{glossaries}
\makeglossaries

\newacronym{ml}{ML}{Maximum Likelihood}
\newacronym{pdf}{PDF}{Probability Density Function}
\newacronym{los}{LoS}{line-of-sight}
\newacronym{nlos}{NLoS}{non-line-of-sight}
\newacronym{leo}{LEO}{Low-Earth Orbit}
\newacronym{rt}{RT}{Ray Tracing}

\begin{document}

\title{A Ray-Based Characterization of Satellite-to-Urban Propagation}


\author{
Nicolò Cenni, Marina Barbiroli, Vittorio Degli-Esposti,\\ Enrico M. Vitucci, Carla Amatetti, Franco Fuschini
\thanks{The authors are with the Department of Electrical, Electronic and Information Engineering “G. Marconi,” CNIT, University of Bologna, Italy, e-mail: {nicolo.cenni2, marina.barbiroli, v.degliesposti, enricomaria.vitucci, carla.amatetti2, franco.fuschini} @unibo.it}
\thanks{This work has been carried out in the framework of the 6G-NTN project, funded by the Smart Networks and Services Joint Undertaking (SNS JU) under the European Union’s Horizon Europe research and innovation programme under Grant Agreement No 101096479.
The work was also supported in part by the EU COST Action INTERACT (Intelligence-Enabling Radio Communications for Seamless Inclusive Interactions), Grant CA20120.
}

}

\maketitle

\begin{abstract}
The evolution toward 6G communication systems is expected to rely on integrated three-dimensional network architectures where terrestrial infrastructures coexist with non-terrestrial stations such as satellites, enabling ubiquitous connectivity and service continuity. In this context, accurate channel models for satellite-to-ground propagation in urban environments are essential, particularly for user equipment located at street level where obstruction and multipath effects are significant. This work investigates satellite-to-urban propagation through deterministic ray-tracing simulations. Three representative urban layouts are considered, namely dense urban, urban, and suburban. Multiple use cases are investigated, including handheld devices, vehicular terminals, and fixed rooftop receivers operating across several frequency bands. The analysis focuses on the relative importance of competing propagation mechanisms and on four key channel parameters, namely shadowing standard deviation, Rician \(K\)-factor, delay spread and Doppler spread, which are relevant for the calibration of channel models to be used in link- and system-level simulations. Results show that elevation and receiver placement strongly shape fading and dispersion, antenna directivity affects shadowing differently under LoS and NLoS conditions, urban morphology produces nontrivial multipath trends, and terminal mobility largely governs Doppler spread.
\end{abstract}

\vskip0.5\baselineskip

\section{Introduction}
The evolution toward 6G networks to achieve ubiquitous and high-capacity global connectivity demands the deployment of unified, seamless 3D network architectures where terrestrial infrastructures and non-terrestrial nodes such as Unmanned Aerial Vehicles, High Altitude Platforms and satellites complement each other~\cite{Unified_3D_networks}. The 3D architecture not only ensures reliable coverage for rural and underserved areas and maintains connectivity during disruptive events such as natural disasters but also enhances the performance of traditional terrestrial networks in a cost-effective way. In addition, it supports service availability under high-mobility conditions and strengthens machine-to-machine communications by enabling seamless service continuity.
This new integrated 3D architecture is expected to include not only line-of-sight (LoS) user equipment (UE), provided with directive antennas on top of buildings, but also pedestrian UE at street level and/or in vehicular mobility in non LoS (NLoS) locations, equipped with omnidirectional antennas. These ground stations can be deployed within dense urban environments, where obstructions such as street furniture and buildings can introduce significant propagation impairments and strongly impact the received signal quality.
In this scenario, the availability of a highly accurate and reliable channel model, accounting for different propagation characteristics (multipath propagation in NLoS conditions at ground level and LoS propagation on top of buildings), is crucial to effectively assess the overall system performance~\cite{Overview_Performance_analysis}. In this respect, performance and challenges of channel models for Non Terrestrial Networks (NTN) are reported in~\cite{Channel_model_performance_analysis} and in~\cite{Survey_NTN}, with particular emphasis on the limited availability of measurement data,  including datasets obtained by emulating satellite links using UAVs or aircraft. The survey further shows that the channel characterization mainly focuses on parameters such as fading statistics, often reported to follow a log-normal distribution ~\cite{Log-normal}, and other channel-state descriptors. Recent studies have investigated satellite-to-ground propagation using statistical~\cite{Statistical_model}, geometry-based stochastic~\cite{Geom_stoch}, deterministic~\cite{Detrministic}, and measurement-based approaches; however, a systematic characterization of multipath parameters for street-level users under both LoS and NLoS conditions, across different urban environments, elevation angles, and frequency bands, remains limited. Among the available studies, 3GPP TR 38.811~\cite{3gppTR38811} represents a key reference for NTN link-level simulations and is therefore adopted in this work. It builds on the 3GPP New Radio channel model defined in TR 38.901~\cite{3gppTR38901} and on ITU-R P.681~\cite{iturP681}, and is specified for the S and Ka bands. The overall framework includes large-scale propagation effects, such as LoS probability, path loss (PL), shadow fading, outdoor-to-indoor (O2I) penetration loss, atmospheric absorption, rain and cloud attenuation, and scintillation, together with fast fading and a time-varying Doppler component. Multipath fading is represented through Tapped Delay Lines (TDL) models, where each tap represents a cluster of multipath components with similar delays and is characterized by statistically modeled amplitudes, typically following a Rician distribution under LoS conditions and a Rayleigh distribution under NLoS conditions. When properly parameterized, TDL models can be adapted to different propagation environments and elevation angles.
Other models rely on the clutter loss (CL) evaluation, defined as the attenuation of signal power caused by the surrounding clutter of the receivers, e.g., buildings and vegetation on the ground~\cite{Multiband_Clutter_loss, 9599387}. In particular, in~\cite{Multiband_Clutter_loss}, some empirical values are provided under different elevation angles for the S-band and the Ka-band.
Unfortunately, studies on Satellite-to-Ground (S2G) propagation in urban environments remain scarce, and although clutter loss and path loss in the earth-space section of the link \cite{Multiband_Clutter_loss, 9599387,5342290, 9134887} have already been studied in detail, a thorough multipath analysis addressing fading characterization as well as channel dispersion is still needed.
Furthermore, existing standard models, including the 3GPP model, still exhibit limitations in both scope and reliability. As highlighted in~\cite{EuCAP2024_satellite}, these include counter-intuitive \(K\)-factor trends in LoS links, the lack of \(K\)-factor values for NLoS conditions, and a frequency range restricted to the S and Ka bands. The lack of NLoS characterization is particularly relevant at low elevation angles and for street-level receivers in urban and dense-urban environments, where blockage of the direct path is more likely to occur.

This paper aims at addressing these gaps by extending the evaluation to the C, Q, and V frequency bands, and providing \(K\)-factor and Delay Spread (DS) estimates in both LoS and NLoS locations. These two parameters are key propagation indicators, as they capture the channel’s dispersive characteristics and are essential inputs for TDL-based link-level simulators. Among small-scale fading effects, particular attention has been also devoted to the Doppler characteristics, since even relatively small frequency shift can lead to noticeable performance degradation in high-frequency broadband satellite systems operating in the Ka- and V-bands~\cite{Survey_NTN}.

Moreover, large-scale channel parameters, including LoS probability and shadow fading, have been evaluated for the considered environments.

The major scope of the study is thus centered on the effects of the urban environment on propagation mechanisms and channel statistics. From a propagation perspective, the key innovative aspect of the 3D architecture envisioned for future satellite systems lies in enabling signal reception at street level and in NLoS conditions. The contribution of this work lies in the more detailed characterization of urban multipath propagation and its dependence on the different use cases envisaged for future NTN.

Absolute path loss and atmospheric propagation effects are not explicitly evaluated, as they are assumed to be separately accounted for using established models~\cite{5342290, 9134887,Survey_NTN}. Similarly, the additional attenuation associated with street-level reception relative to rooftop reception is already addressed by existing 3GPP models~\cite{3gppTR38811,3gppTR38901}.

The analysis is carried out through an image-based 3D \gls{rt} tool, developed in house at the University of Bologna and widely validated vs. measurements in urban environment~\cite{Ray_Tracing_UniBO}, capable of simulating multipath propagation in indoor and outdoor environments with multiple interactions, including specular reflection, transmission, diffraction, diffuse scattering and combinations of these. Simulations are carried out in three different urban scenarios, characterized by different building densities, namely: dense urban, urban and suburban.
It is worth noting that the main objective of the present study is to analyze the impact of propagation within the urban structure in the final segment of the satellite-to-urban link and to improve its modeling.

The remainder of the paper is organized as follows: Section II describes the simulation framework, including the geometrical description of the scenarios, the link and communication parameters, the ray-tracing simulator, the ray-geometry reuse and narrowband-array approximation for multi-frequency and multi-antenna evaluation, and the computation of the target channel parameters. Section III presents and discusses the simulation results, while Section IV draws the main conclusions.

\section{Simulation Framework}

\subsection{Geometrical Description of the Scenarios}

\begin{figure}[]
    \centering
    \begin{subfigure}[b]{0.51\linewidth}
        \centering
        \includegraphics[width=\linewidth]{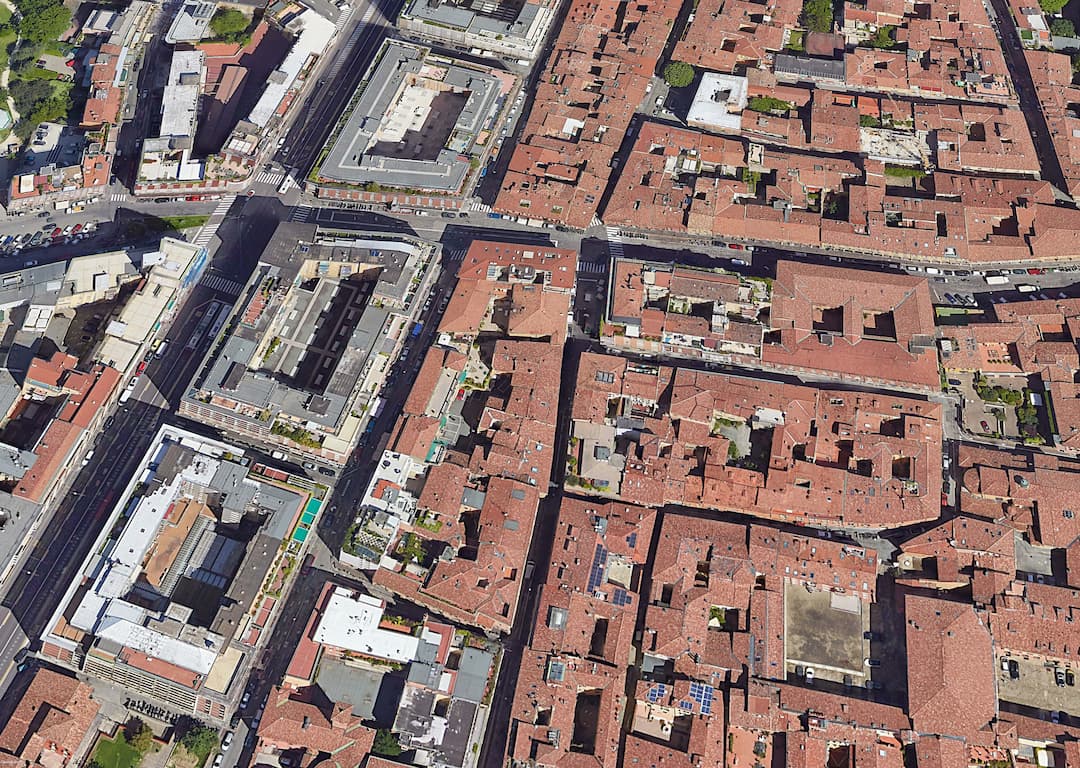}
        \caption{}
        \label{fig:denseurban1}
    \end{subfigure}
    \hfill
    \begin{subfigure}[b]{0.47\linewidth}
        \centering
        \includegraphics[width=\linewidth]{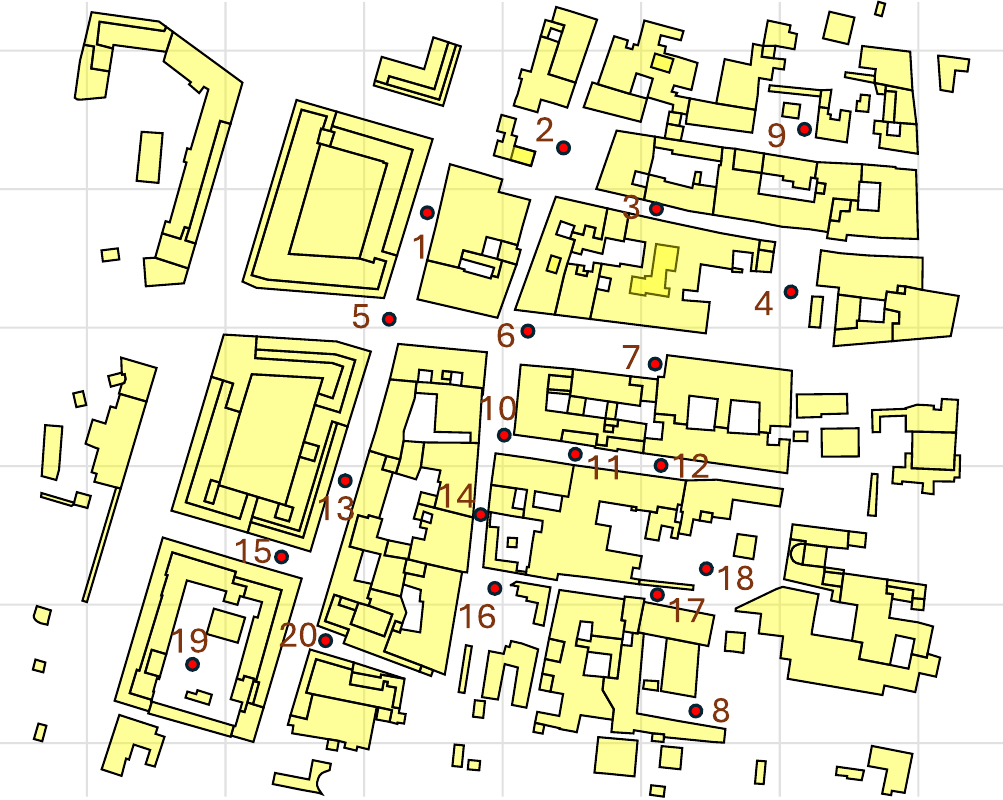}
        \caption{}
        \label{fig:denseurban2}
    \end{subfigure}
    \caption{Bologna dense urban: aerial overview (a) and digital map for simulations highlighting the 20 receiver grids’ positions (b).}
    \label{fig:denseurban}
\end{figure}

\begin{figure}[]
    \centering
    \begin{subfigure}[b]{0.54\linewidth}
        \centering
        \includegraphics[width=\linewidth]{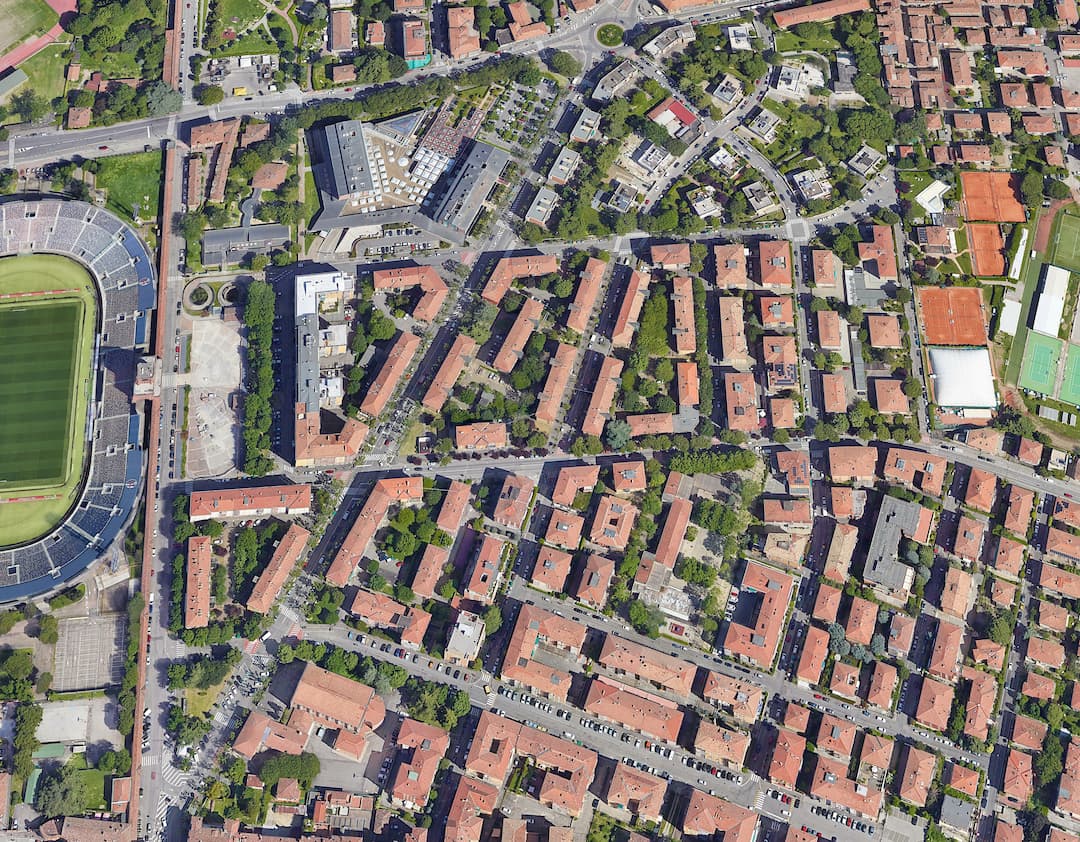}
        \caption{}
        \label{fig:urban1}
    \end{subfigure}
    \hfill
    \begin{subfigure}[b]{0.44\linewidth}
        \centering
        \includegraphics[width=\linewidth]{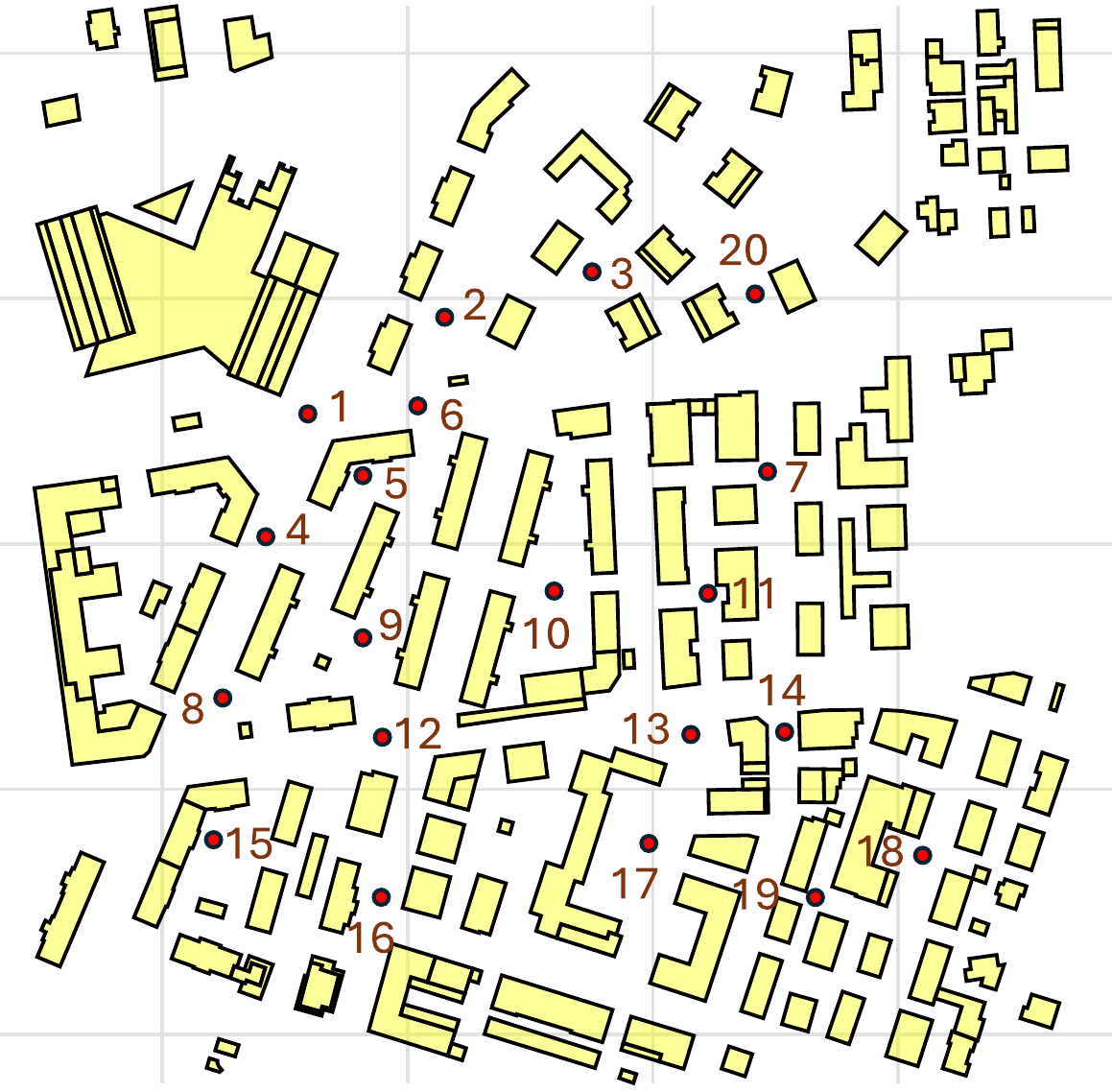}
        \caption{}
        \label{fig:urban2}
    \end{subfigure}
    \caption{Bologna urban: aerial overview (a) and digital map for simulations highlighting the 20 receiver grids’ positions (b).}
    \label{fig:urban}
\end{figure}

\begin{figure}[]
    \centering
    \begin{subfigure}[b]{0.48\linewidth}
        \centering
        \includegraphics[width=\linewidth]{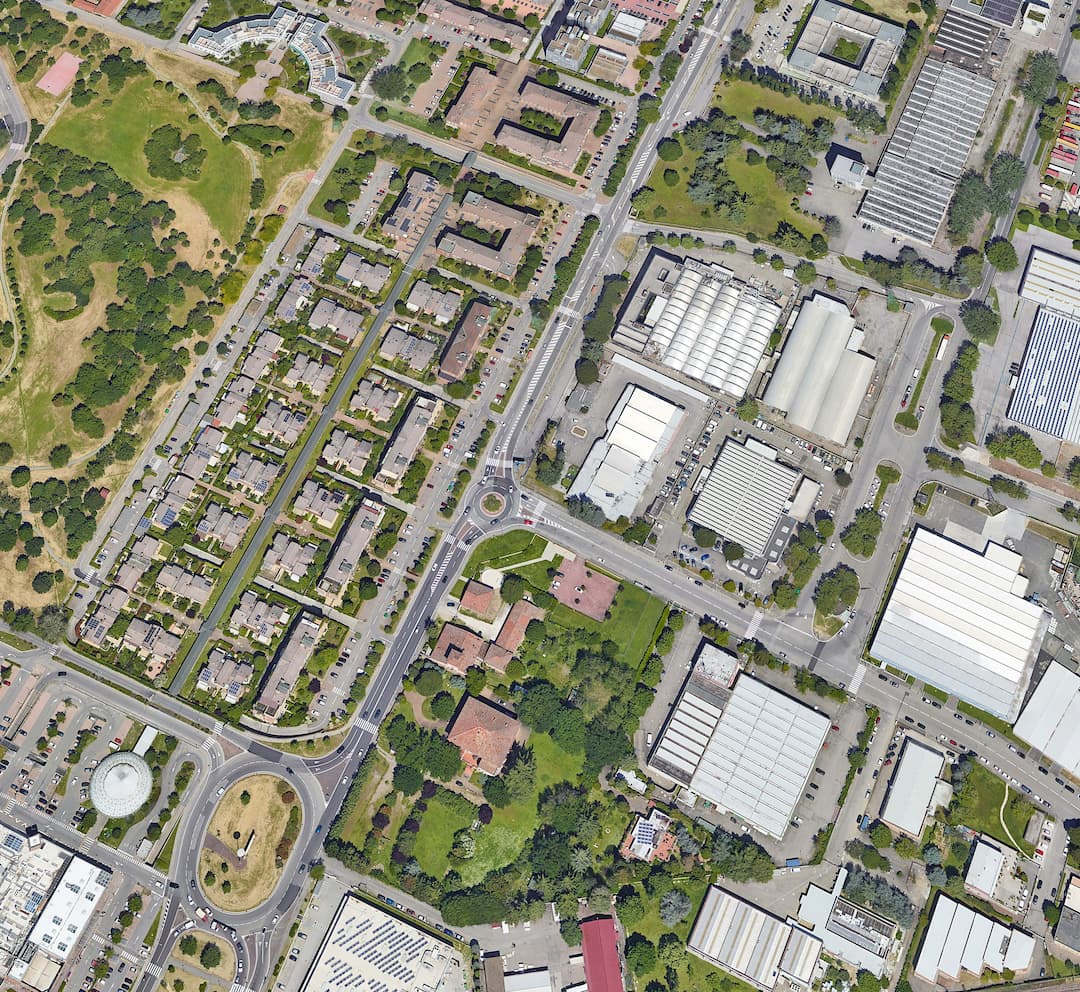}
        \caption{}
        \label{fig:suburban1}
    \end{subfigure}
    \hfill
    \begin{subfigure}[b]{0.50\linewidth}
        \centering
        \includegraphics[width=\linewidth]{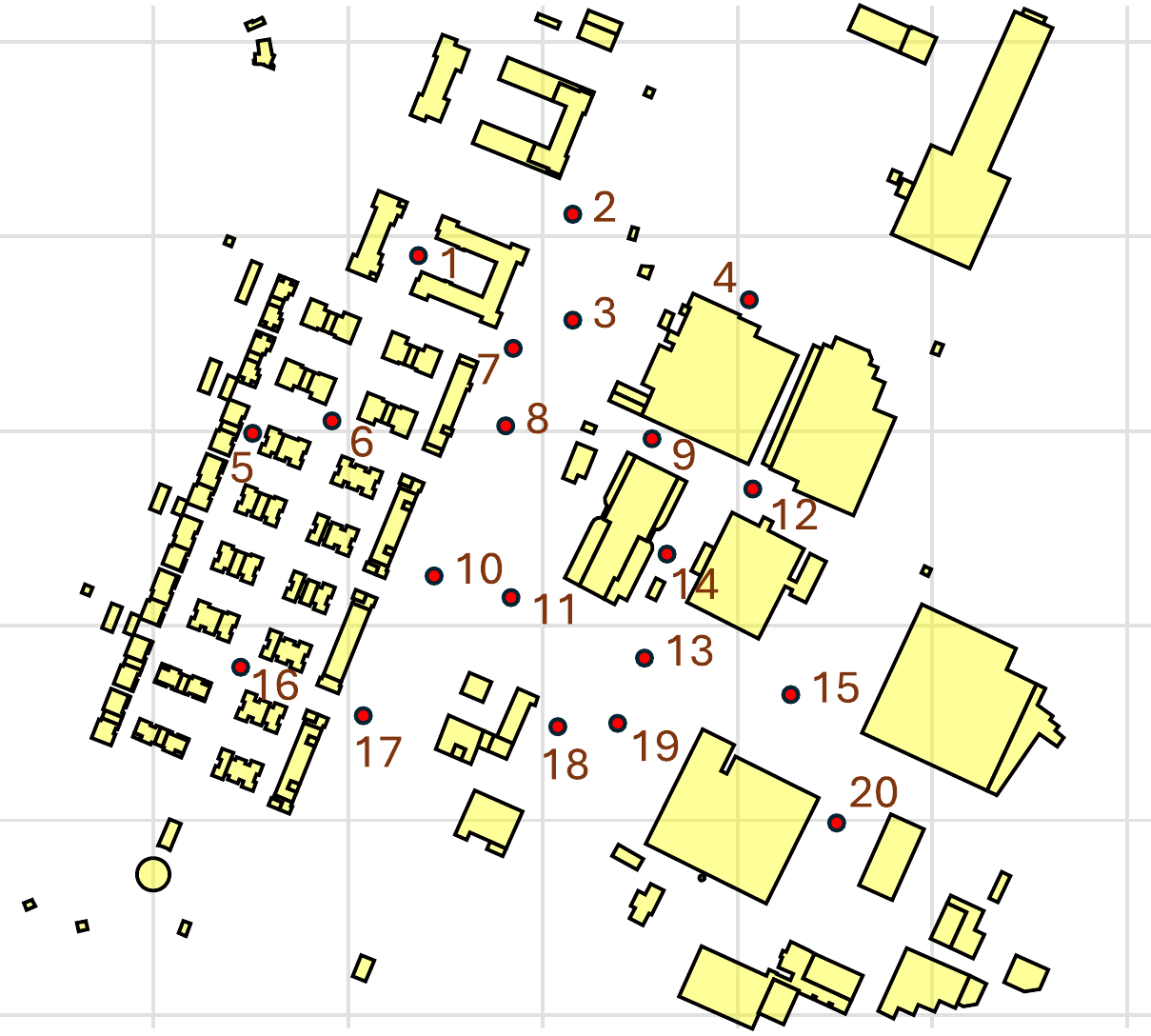}
        \caption{}
        \label{fig:suburban2}
    \end{subfigure}
    \caption{Bologna suburban: aerial overview (a) and digital map for simulations highlighting the 20 receiver grids’ positions (b).}
    \label{fig:suburban}
\end{figure}

Three areas of the city of Bologna (Italy) were selected as reference scenarios for the simulations and were reconstructed based on a digital 3D vector database, with buildings, streets and other urban elements modeled as solid prisms, i.e. 2D polygons with height. The dense urban scenario (Fig.~\ref{fig:denseurban}) corresponds to the historical city center, characterized by narrow streets, enclosed courtyards, and a dense grid of mid-rise buildings. It covers an area of \(340\,\text{m} \times 290\,\text{m}\) and includes 200 buildings, with a mean height of \(15\,\text{m}\) and a \(9\,\text{m}\) standard deviation (Table \ref{tab:geometrical_data}). The average street width is approximately equal to \(9.3\,\text{m}\) (Table \ref{tab:geometrical_data}). The complex structure, including both small alleys and open squares, leads to frequent LoS obstructions and a high level of multipath components due to reflections and diffuse scattering.
The urban scenario is represented by an area close to the city Stadium, a typical residential neighborhood featuring medium-density housing and small private gardens (Fig.~\ref{fig:urban}). The area spans \(440\,\text{m} \times 440\,\text{m}\) and includes 188 buildings, with a mean height of approximately \(13\,\text{m}\) with a standard deviation of \(7\,\text{m}\), and an average street width equal to about \(17\,\text{m}\) (Table \ref{tab:geometrical_data}). The geometry here provides a mixture of LoS and NLoS conditions, depending on the elevation angle of the satellite. Compared to the dense urban area, this environment has more open spaces, resulting in improved LoS availability, especially at moderate to high elevation angles.
The suburban scenario encompasses a \(540\,\text{m} \times 540\,\text{m}\) peripheral district of Bologna where low-rise residential buildings are interspersed with industrial structures and green areas (Fig.~\ref{fig:suburban}). This configuration results in significantly lower building density, corresponding to an increased average street width up to \(30\,\text{m}\). The average building height is \(8\,\text{m}\), with a standard deviation of \(5\,\text{m}\), for a total of 184 buildings (Table \ref{tab:geometrical_data}). Due to the sparse layout and reduced obstructions, this environment offers the highest LoS probability, even at low elevation angles, and typically shows lower multipath richness.

In order to perform a statistically significant analysis, 20 square receiver grids, each measuring \(4\,\text{m}\) per side and containing \(15 \times 15\) receivers, are strategically placed throughout the map in each scenario, to capture as much environmental variability as possible. An overview of the selected receiver spots in the digital maps is provided in Figs.~\ref{fig:denseurban2},~\ref{fig:urban2},~\ref{fig:suburban2}. The grid spacing of approximately $29\;cm$ ensures that spatial samples can be considered uncorrelated (or at least weakly correlated) across all frequency bands considered in this study, given that—in the presence of Rician fading and a sufficiently wide angular distribution of scatterers, as is the case in the urban environments under investigation—the correlation distance rarely exceeds a few wavelengths \cite{spatial_correlation}.

\begin{figure}[]
    \centering
    \begin{subfigure}[b]{0.54\linewidth}
        \centering
        \includegraphics[width=\linewidth]{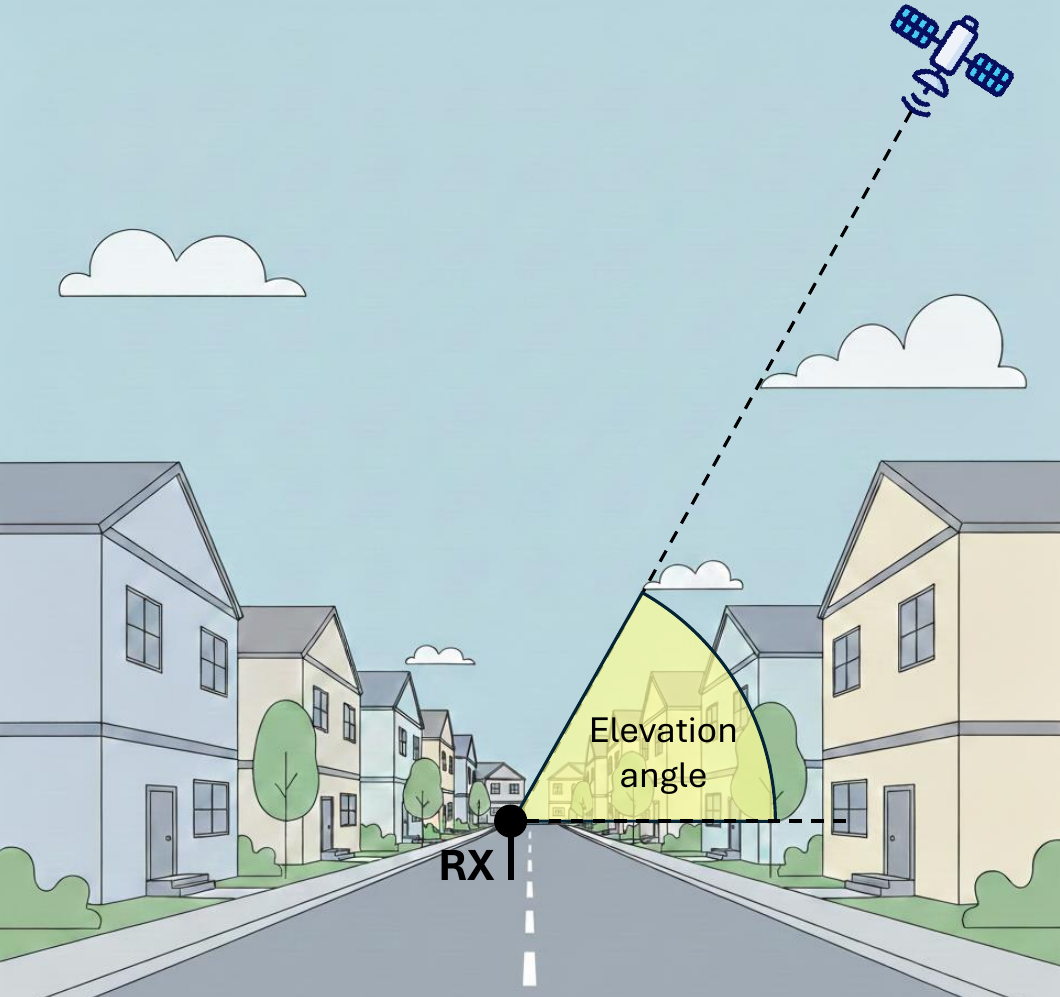}
        \caption{}
        \label{fig:elevation}
    \end{subfigure}
    \hfill
    \begin{subfigure}[b]{0.44\linewidth}
        \centering
        \includegraphics[width=\linewidth]{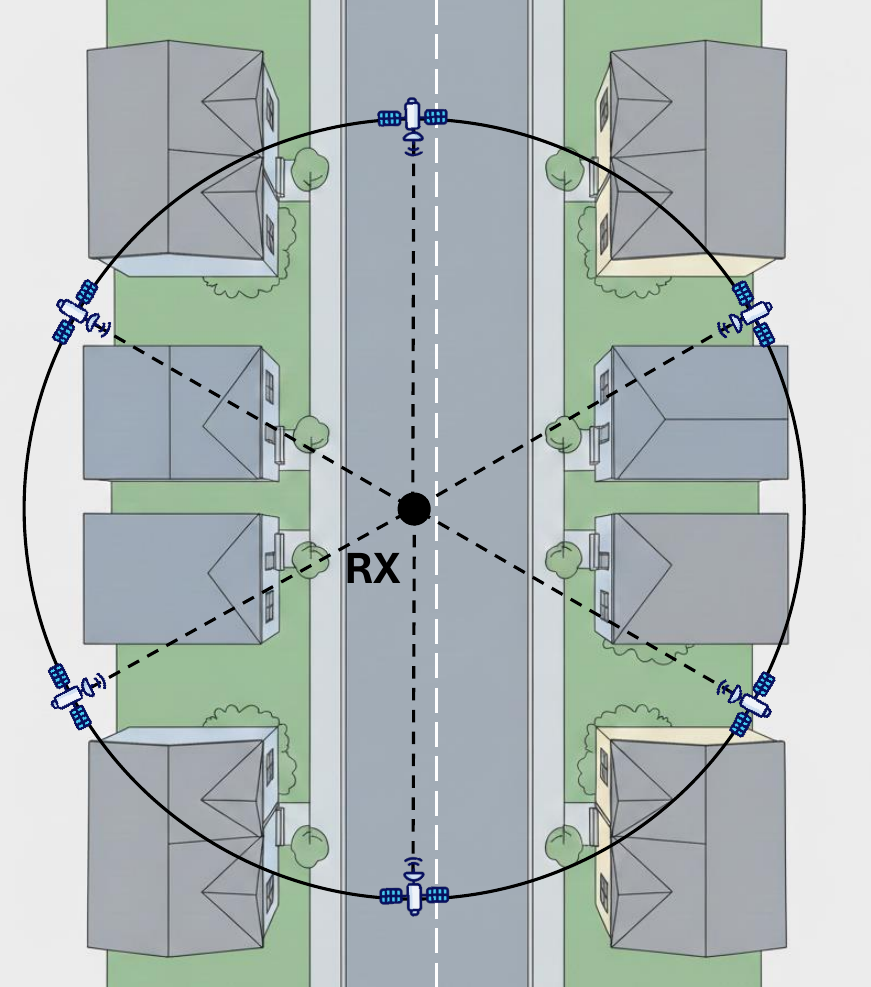}
        \caption{}
        \label{fig:azimuth}
    \end{subfigure}
    \caption{(a) Side view illustrating the satellite elevation angle. (b) Top view showing an example configuration of six satellite azimuth angles uniformly spaced by 60°. Image generated using AI (Google Gemini) and refined by the authors.}
    \label{fig:satellite_angles}
\end{figure}

The aerial transmitting station simulates a \gls{leo} satellite, assumed to be located at a distance of \(500\,\text{km}\) from the Earth surface. For each grid, nine satellite elevation angles were considered (Fig.~\ref{fig:elevation}), ranging from \(10^\circ\) to \(90^\circ\) in \(10^\circ\) increments. In addition, six azimuth angles were selected, starting from a randomly chosen direction and uniformly spaced at \(60^\circ\) intervals (Fig.~\ref{fig:azimuth}).

The main information related to the geometrical description of the scenarios is summarized in Table \ref{tab:geometrical_data}.

\begin{table}[h!]
\centering
\caption{Simulations Geometrical Parameters}
\resizebox{\columnwidth}{!}{
\begin{tabular}{|l|c|c|c|}
\hline
 & \textbf{Dense Urban} & \textbf{Urban} & \textbf{Suburban} \\
\hline
Map size & 340 m $\times$ 290 m & 440 m $\times$ 440 m & 540 m $\times$ 540 m \\
\hline
No. of buildings & 200 & 188 & 184 \\
\hline
Building density (no./km$^{2}$) & 2028 & 971 & 631 \\
\hline
Mean building height & 15 m & 13 m & 8 m \\
\hline
Average street width & 9.3 m & 17 m & 30 m \\
\hline
Elevation range & \multicolumn{3}{c|}{10$^\circ$, 20$^\circ$, 30$^\circ$, 40$^\circ$, 50$^\circ$, 60$^\circ$, 70$^\circ$, 80$^\circ$, 90$^\circ$} \\
\hline
Grid size & \multicolumn{3}{c|}{4 m} \\
\hline
No. of receivers per grid & \multicolumn{3}{c|}{225} \\
\hline
Space to earth distance & \multicolumn{3}{c|}{500 km} \\
\hline
\end{tabular}
}
\label{tab:geometrical_data}
\end{table}

\subsection{Link and Communication Parameters}

Since the receiving locations can be of course assumed well included inside the footprint of the satellite antenna in real applications, an isotropic radiation pattern has been considered at the aerial station for the sake of simplicity, with circular polarization in agreement with \cite{3gppTR38811}. Three different antenna types are instead considered for the Earth equipment, depending on the communication frequency and addressing different use cases \cite{3gppTR38811}:
\begin{itemize}
    \item \textit{Isotropic}: it accounts for \textit{handheld} (or IoT) devices, working in the lowest frequency band (S/C) at 1.5 m from the ground level;
    \item \textit{Patch-like}: it still works in the S/C bands and is conceived for vehicular application, where the antenna is placed on car roofs, with a mild directivity and vertical boresight direction regardless of the satellite position;
    \item \textit{60 cm equivalent aperture} (V-SAT in \cite{3gppTR38811}): it represents a directive antenna intended for fixed reception. In this case, to achieve a higher probability of LoS, the user equipment is raised at the (average) rooftop level, with antenna boresight always pointed towards the satellite position. This antenna is most suited to the higher frequency bands, i.e the Ka, Q and V bands, where a reasonably sized antenna can still provide adequate performance.
\end{itemize}
The main properties of the different considered antennas at the earth end user are summarized in Table \ref{tab:use_cases}.

\begin{table*}[h!]
\centering
\caption{Use cases}
\begin{tabular}{|c|c|c|c|c|c|c|}
\hline
\textbf{Use Case} & \textbf{Earth Antenna Type} & \textbf{Polarization} & \textbf{Frequency Band} & \textbf{Directivity} & \textbf{HPBW} & \textbf{Position} \\
\hline
Handheld & Isotropic & \multirow{2}{*}{Linear (vertical)} & \multirow{2}{*}{S (1.98-2.2 GHz)} & 0 dBi & -- & Ground level (1.5 m) \\
\cline{1-2}\cline{5-7}
Vehicular & Patch-like &  & / C (3.4-3.9 GHz) & $\sim 6$ dBi & \makecell{120$^\circ$ \\ (both E- and H-planes)} & Car Roof (1.5 m) \\
\hline
\multirow{3}{*}{Fixed} &
\multirow{3}{*}{60 cm Aperture} &
\multirow{3}{*}{Circular} &
Ka (17-30 GHz) & $\sim 44$ dBi & 2.9$^\circ$ & \multirow{3}{*}{Average Building Height} \\
\cline{4-6}
 &  &  & Q (36-46 GHz) & $\sim 48$ dBi & 1.8$^\circ$ & \\
\cline{4-6}
 &  &  & V (46-56 GHz) & $\sim 50$ dBi & 1.5$^\circ$ & \\
\hline
\end{tabular}
\label{tab:use_cases}
\end{table*}

\subsection{Ray Tracing Simulator}
The simulations are conducted using the 3D \gls{rt} tool developed at the University of Bologna~\cite{Ray_Tracing_UniBO}.
The input data of the \gls{rt} simulator consists of "Shape" format databases, where buildings and objects are modeled as polygons with height, corresponding to the rooftop level with respect to ground. The electromagnetic parameters (relative permittivity $\epsilon_r$ and electric conductivity $\sigma$) for each building wall are also included, as well as the 3D radiation characteristics of the antennas including the field polarization.
Specular propagation mechanisms (reflections, edge diffractions) are modeled using geometric optics theory and its extensions, like the uniform theory of diffraction \cite{mcnamara1990introduction,UTD1974,UTD_heuristic}.
The field carried by the m-th ray at a given receiver location is computed according to the following equation:

\begin{equation}
\begin{gathered}
  {\mathbf{E}}^m(Rx) = \left[ {\prod\limits_{k = 0}^{{N_m}} {{{\underline {\mathbf{D}} }_k}\left(\lambda_0,{\varepsilon _r}, \sigma \right) \cdot {\mathbf{E}}_{Tx_0}} } \right] \cdot A\left( {{r_0},{r_1},...,{r_{{N_m}}}} \right)  \hfill \\
  \begin{array}{*{20}{c}}
  {\begin{array}{*{20}{c}}
  {}&{}
\end{array}}&{ \cdot {\text{ }}{e^{ - j\frac{{2\pi }}{{{\lambda _0}}}\left( {{r_0} + {r_1} + ... + {r_{{N_m}}}} \right)}}}
\end{array} \hfill \\
\end{gathered}
\label{coherent_field}
\end{equation}

where $N_m$ is the number of bounces of the m-th ray, ${{\mathbf{E}}_{Tx_0}}={\mathbf{E}}_{Tx}\left(r=1,    \theta_0,\phi_0 \right)$ is the electric field emitted by the Tx antenna toward the direction of the first reflection/diffraction point -- or toward the Rx, in the case of LoS rays -- at a reference distance of $1\,\mathrm{m}$, $\lambda_0=c_0/f$ is the free-space wavelength corresponding to the simulation frequency, $r_k$ with $k=0,1,..,N_m$ is the length of the k-th ray segment,
${{\underline {\mathbf{D}} }_k}$ is a dyadic interaction coefficient that depends on the k-th interaction type (reflection or diffraction), as well as the wavelength and the electromagnetic parameters,
$A$ is the \emph{spreading factor} that takes into account the total attenuation along the ray, and the complex exponential takes into account the phase shift corresponding to the total ray length.

For LoS rays, the multiplication of dyadics in \eqref{coherent_field} degenerates into the identity tensor $\underline {\mathbf{D}}_0=\underline {\mathbf{1}}$, while the attenuation factor boils down to $A=1/r_0$.
For rays with specular reflections only, the dyadics $\underline {\mathbf{D}}_k$ are computed through the well-known Fresnel's coefficients, while the attenuation factor becomes $A=(r_0+r_1+..+r_{N_m})^{-1}$. The expressions of dyadic coefficients and attenuation that are used for rays with edge diffractions can be found in \cite{mcnamara1990introduction, UTD1974}.

In addition to specular interactions, one key feature of the simulator is the integration of a diffuse scattering model based on the Effective Roughness (ER) approach, which takes into account non-specular diffuse scattering mechanisms caused by surface or volume irregularities \cite{scat2007,scat2023, 6183484}.
The basic assumption of the ER model is that irregular walls are divided into elements, or "tiles" of a given area $\Delta A$, each generating a scattered field computed as \cite{scat2023,6183484}:
\begin{equation}
\mathbf{E}_s(Rx) = K_{T}\;\frac{S\:\Gamma({{\mathbf{\hat k}}}_i)}{{{r_i}\:{r_s}}} \cdot \sqrt{\frac{f({{\mathbf{\hat k}}}_i,{{\mathbf{\hat k}_s}})}{F({{\mathbf{\hat k}}}_i)} \: \Delta A \cos\vartheta_i } \cdot {\mathbf{\hat{p}}_s}\;{e^{j\chi_s }}
\label{scattering_field}
\end{equation}
where ${{\mathbf{\hat k}}}_i,{{\mathbf{\hat k}_s}}$ are the incidence and observation directions, respectively, $K_T$ is an amplitude function depending on the Tx power and Tx antenna gain in the incidence direction ${{\mathbf{\hat k}}}_i$, $S\in \left[ {0,1} \right]$ is the \textit{scattering parameter} which depends on the degree of irregularity of the wall and expresses the percentage of field intensity diffused in non-specular directions at the expense of specular reflection, $\Gamma({{\mathbf{\hat k}}}_i)$ is the specular reflection coefficient computed for incidence direction ${{\mathbf{\hat k}}}_i$,  $f({{\mathbf{\hat k}}}_i,{{\mathbf{\hat k}_s}})$ is a function that represents the shape of the scattering pattern, $\vartheta_i=\arccos{(-{{\mathbf{\hat k}}}_i \cdot {\mathbf{\hat n}})}$ is the incidence angle, with ${\mathbf{\hat n}}$ standing for the unit vector orthogonal to the wall surface, $r_i$ and $r_s$ are the length of the ray segments connecting the tile center with Tx and Rx, respectively, $\mathbf{\hat{p}}_s$ is the polarization vector of the scattered field, $\chi_s \in \left[ {-\pi,\pi} \right]$ is a uniformly distributed random variable representing the phase of the scattering path, and $F$ is a proper normalization factor of the scattering pattern, obtained through integration of the scattering function $f$ on the whole half-space in front of the wall.

The scattering function $f({{\mathbf{\hat k}}}_i,{{\mathbf{\hat k}}}_s)$ can have different forms, but the one that gave the best results in urban environments for microwave and mm-wave frequencies -- and therefore used in the present work -- is the single-lobe directive pattern, centered on the direction of specular reflection \cite{scat2007,scat_pam_1,scat_pam_2}:
\begin{equation}
f({{\mathbf{\hat k}}}_i,{{\mathbf{\hat k}}}_s)={\left( {\frac{{1 + \cos {\psi _R}}}{2}}\right)^{{\alpha _R}}}
\end{equation}
where the exponent $\alpha_R$ is a parameter that sets the pattern directivity -- the higher $\alpha_R$, the narrower the lobe -- while $\psi_R$ is the \textit{off-specular angle}, that satisfies the following relation:
\[
\cos\psi_R={{{\mathbf{\hat k}}}_r \cdot {{\mathbf{\hat k}}}_s}=\left[ {{{{\mathbf{\hat k}}}_i} - 2( {{{{\mathbf{\hat k}}}_i} \cdot {\mathbf{\hat n}}}){\mathbf{\hat n}}} \right] \cdot {\mathbf{\hat k}_s}
\]

In order to ensure physical consistency of the ER model, a power balance between non-specular diffuse scattering and specular interaction mechanisms -- reflection, diffraction -- is enforced. This leads to the relation $S^2+R^2=1$ with $R$ being the "specular reduction factor", as detailed in \cite{scat2007}.
Correspondingly, the reduction factor $R=\sqrt{1-S^2}$ is applied in \eqref{coherent_field} for all specular rays bouncing on walls where diffuse scattering is enabled (i.e. with $S>0$).

It is worth noting that, in order to calculate the rician \(K\)-factor, which is a propagation parameter dependent on multipath fading characteristics, an  evaluation of the "coherent" field is necessary, including its polarization and phase properties. This is naturally done for LoS and rays with specular interactions (reflections, edge diffractions) in accordance with \eqref{coherent_field}.
For diffuse scattering, polarization is modeled in \eqref{scattering_field} through the polarization vector $\mathbf{\hat{p}}_s$ which takes into account polarization rotations introduced by the scattering interaction, while the random phase $\chi_s$ takes into account the stochastic nature of the diffuse scattering mechanisms, that can be interpreted as the macroscopic effect of multiple micro-interactions (mainly diffractions) caused by the rough surface and the internal volume irregularities \cite{6183484}.

Sequential combinations of different types of mechanisms such as reflections and diffractions in \eqref{coherent_field} are also possible, as well as reflections/diffractions combined with diffuse scattering by cascading \eqref{coherent_field} and \eqref{scattering_field}.

The maximum number of interactions considered in \gls{rt} simulations for different interaction types and their combinations is reported in Table \ref{tab:ray_tracing_parameters1}.

The electromagnetic parameters used in the simulations with reference to different frequency bands and scenarios -- including the relative dielectric permittivity, conductivity, and the ER-model scattering parameter \(S\) -- were obtained from the literature and previous work \cite{scat_pam_1,scat_pam_2,11075837, 875221}. Values are listed in Table \ref{tab:ray_tracing_parameters2} for building walls, and in Table \ref{tab:ray_tracing_parameters3} for the terrain.
The foregoing material parameter values represent typical, mean permittivity and conductivity values for European street pavements and building walls, which are commonly composed of asphalt, reinforced concrete, bricks, and plaster. These parameters should not be interpreted as corresponding exclusively to one of those materials, but rather as representative effective values for typical construction materials. Ray tracing predictions have been shown to be relatively insensitive to moderate variations in material parameters  \cite{875221}.

The choice of diffuse scattering parameters for the terrain was made considering that a significant portion of the scattering contribution from the ground depends on the fraction of area occupied by vehicles, which has been estimated through multiple representative maps for each type of environment.
Therefore, the \(S\) values in Table \ref{tab:ray_tracing_parameters3} are highest for the dense urban scenario, characterized by the greatest vehicle density, followed by the urban and suburban scenarios. Moreover, the $S$ values increase with frequency, as the materials surface roughness becomes increasingly more significant at shorter wavelengths.

Finally, in accordance with the previously cited works, the scattering tile area $\Delta A$ was kept fixed to $5\times5\;m^2$, while the parameter $\alpha_{R}$ was set for all simulations equal to $2$, which proved to be a suitable value for urban environments\cite{scat_pam_1}.

\begin{table}[!t]
\centering
\label{tab:max_interactions}
\caption{Maximum number of interactions for each mechanism considered in \gls{rt} simulations. The same values are adopted for all frequency bands and scenarios}
\begin{tabular}{lc}
\hline
\textbf{Interaction Mechanism} & \textbf{Max. No. of Interactions} \\
\hline
\# Reflections alone                      & 3 \\
\# Diffractions alone                     & 2 \\
\# Scatterings alone                      & 1 \\
\# Reflections \& scatterings together    & 2 \\
\# Reflections \& diffractions together   & 2 \\
\# Diffractions \& scatterings together   & 0 \\

\hline
\end{tabular}
\label{tab:ray_tracing_parameters1}
\end{table}
\begin{table}[t]
\centering
\caption{Electromagnetic parameters used in \gls{rt} simulations for buildings walls at two frequency bands. Identical values are assumed for all scenarios}
\begin{tabular}{lcc}
\hline
\textbf{Parameters}
& \textbf{S/C bands}
& \textbf{Ka/Q/V bands} \\
\hline
Relative dielectric permittivity $\varepsilon_r$
& 5 & 5.5 \\
Conductivity (S/m)
& 0.01 & 0.4 \\
Diffuse scattering parameter $S$
& 0.4 & 0.6 \\
\hline
\end{tabular}
\label{tab:ray_tracing_parameters2}
\end{table}

\begin{table*}[t]
\centering
\caption{Electromagnetic parameters of terrain used in \gls{rt} simulations for the 3 scenarios and the 2 frequency bands under study}
\begin{tabular}{lcccccc}
\hline
\textbf{Parameters}
& \multicolumn{3}{c}{\textbf{S/C bands}}
& \multicolumn{3}{c}{\textbf{Ka/Q/V bands}} \\
\cline{2-4}\cline{5-7}
& Dense urban & Urban & Suburban
& Dense urban & Urban & Suburban \\
\hline
Relative dielectric permittivity $\varepsilon_r$
& 5  & 5 &  5
& 5.5  & 5.5 & 5.5  \\
Conductivity (S/m)
& 0.01  & 0.01 & 0.01
&  0.4 & 0.4 &  0.4 \\
Diffuse scattering parameter $S$
& 0.5 & 0.4 & 0.25
& 0.75 & 0.6 & 0.375 \\
\hline
\end{tabular}
\label{tab:ray_tracing_parameters3}
\end{table*}

\subsection{Ray Geometry Reuse and Narrowband Array Approximation for Multi-Frequency and Multi-Antenna Evaluation}

To reduce the computational burden associated with repeated \gls{rt} simulations (e.g. at different frequencies, and/or with different antenna patterns), a methodology was adopted where the geometric, electromagnetic, and spatial aspects of the problem are treated separately.

First, the ray-path geometry is computed only once. This relies on the assumption \textendash{} widely demonstrated in the literature \cite{Kwon2014MultifrequencyRayTracing,Sionna0192RayTracing, Learning_radio} \textendash{} that the set of propagation paths is frequency-independent, as it is determined exclusively by the physical environment and by the relative positions of the transmitter and receiver.
This assumption has recently led to the development of differentiable ray-tracing techniques that enable the use of computationally efficient calibration methods for ray-tracing predictions \cite{Learning_radio}.

Frequency-dependent effects -- such as material electromagnetic properties and phase rotation during propagation -- are introduced in a post processing stage, where the field amplitude and phase of each ray are recalculated according to \eqref{coherent_field} and \eqref{scattering_field} for each frequency of interest.

Second, the same \gls{rt} realization obtained with isotropic antennas is reused to evaluate different receiving antennas operating in different bands. The effect of the actual UE antenna is incorporated by weighting each ray with the corresponding antenna gain in its direction of arrival. Since modifying the radiation pattern does not alter the ray geometry, this approach avoids the need for additional \gls{rt} simulations while preserving deterministic propagation characteristics.

Third, in order to evaluate a dense grid of receiver locations, a "narrowband array" approximation is applied \cite{Steinbauer2001,951559}. The electromagnetic field associated with each ray is computed only at the grid center and then extended to the entire $15 \times 15$ receiver grid by assuming an impinging planar wavefront, thus introducing only phase shifts consistent with the relative receiver positions.
In other terms, the fields computed through \eqref{coherent_field} and \eqref{scattering_field} are multiplied by the phase factor
\begin{equation}
\exp\left[-j\frac{2 \pi}{\lambda_0} \mathbf{\hat{k}}_i \cdot \mathbf{l}_n \right],\;\;\;\;\;\;\;\mathbf{l}_n=\mathbf{r}_n-\mathbf{r}_O
\end{equation}
with the unit vector $\mathbf{\hat{k}}_i$ being the direction of incidence, while $\mathbf{l}_n$ is the vector distance between the n-th location on the Rx grid ($\mathbf{r}_n$) and the grid center ($\mathbf{r}_O$).

Overall, the adopted approach enables multi-frequency, multi-antenna, and multi-location evaluation from a single  \gls{rt} run, providing a substantial reduction in computational complexity.

The methodology adopted in this work is, in broad terms, analogous to the “Spatial filter for generating TDL channel model” described in Section 7.7.4 of~\cite{3gppTR38901}, as both rely on a superposition-based approach. However, it should be noted that the underlying channel models are inherently different: the \gls{rt} is a deterministic point-to-point model, which is able to trace the set of possible paths between the Tx and Rx, while the TDL channel model is a statistical channel model able to represent a multipath fading environment by modeling the impulse response of the channel as a sum of delayed and scaled versions of the transmitted signal.

\subsection{Computation of Target Channel Parameters}
This section outlines how the target propagation parameters, namely the shadowing standard deviation $\sigma_{dB}$, the Rician \(K\)-factor, the delay spread DS and the Doppler spread $B_D$, were evaluated by means of the \gls{rt} simulations.
\subsubsection{Shadowing}
since it refers to large scale fading only, the first step towards its evaluation is the removal of fast fading contributions from \gls{rt} simulations results. The most common way to do this is to compute small-area averages, i.e., to average the received signal levels over each of the twenty ground-level grids \cite{Bertoni}. The spatially averaged power values are then represented vs. link distance in a log-log plot. The shadowing parameter ($\sigma_{dB}$ in the following) is finally computed as the root mean square of the received-power residuals (in dB) relative to the best-fit regression line \cite{Bertoni}. In order to increase the size of the sample set, small area averages corresponding to the six different azimuth values have been merged in the same log-log-plot. For the same reason, values corresponding to the lower elevation angles (10$^\circ$-30$^\circ$) have also been grouped together, as well as values for the middle (40$^\circ$-60$^\circ$) and for the higher (70$^\circ$-90$^\circ$) elevation angles. This is the reason why values of $\sigma_{dB}$ reported in sub-section III.C refer to \textit{low}, \textit{medium} and \textit{high} elevation.

\subsubsection{Rician \(K\)-factor}
it is often defined as the ratio of the power in the dominant component to the power in the remaining scattered components. However, in our scenario -- where channel responses are obtained from a \gls{rt} simulator -- this definition can be ambiguous, as the distinction between the dominant path(s) and the scattered multipath components is not always well-defined. To avoid this ambiguity, we adopt a statistical approach based on the Rician \gls{pdf}. Specifically, we assume that the signal amplitude samples collected within each receiver grid and normalized with respect to their average value can be modeled as \( N \) independent observations of a random variable \( X \) following a Rician distribution with parameters \( \nu \) and \( \sigma \). If \(x_{1},x_{2},\ldots,x_{N}\) represent the observed samples, the Likelihood Function can be written as follows:
\begin{equation}
    L(\nu, \sigma) =\prod_{i=1}^{N} \left [\frac{x_i}{\sigma^2} \exp\left(-\frac{x_{i}^2 + \nu^2}{2\sigma^2}\right) I_0\left(\frac{x_i \nu}{\sigma^2}\right) \right ]
\end{equation}
Here, \( I_0(\cdot) \) denotes the modified Bessel function of the first kind and order zero. Then we estimate the parameters using the \gls{ml} method \cite{vantreesestimation}:
\begin{equation}
    \left ( \hat{\nu},\hat{\sigma}  \right )=\mathrm{argmax}_{\left ( \nu ,\sigma  \right)} L(\nu, \sigma)
\end{equation}
Eventually, the Rician \(K\)-factor estimated value (\(\hat{K}\)) is defined as follows:
\begin{equation}
\hat{K} = \frac{\hat{\nu}^2}{2\hat{\sigma}^2}
\end{equation}
This method provides a consistent and unambiguous characterization of the fading environment, based on the parameters of the underlying statistical model rather than heuristic power separation.
A time efficient alternative to our \gls{ml}-based estimator is the moment‐method proposed by Greenstein et al.~\cite{greenstein1999moment}, which computes the Rician \(K\)-factor from the first and second moments of the received power. Comparing the two methods, the results have shown good agreement.



\subsubsection{Delay Spread}
it describes the temporal dispersion of the radio channel. In this work, the DS is computed only for the receiver located at the central position at each receiver grid for the sake of simplicity. The DS is defined as the square root of the second central moment of the power delay profile and is given by:
\begin{equation}
\mathrm{DS} = \sqrt{\frac{\sum_{i} P_i (\tau_i - \bar{\tau})^2}{\sum_{i} P_i}},
\label{eq:DS}
\end{equation}
where $P_i$ and $\tau_i$ denote the power and delay of the $i$-th multipath component, respectively, and $\bar{\tau}$ is the mean excess delay, defined as
\begin{equation}
\bar{\tau} = \frac{\sum_{i} P_i \tau_i}{\sum_{i} P_i}.
\label{eq:DS_meandelay}
\end{equation}
Both the power carried by each ray ($P_i$) and its propagation delay ($\tau_i$) are direct outputs of the \gls{rt} simulations. It is worth pointing out that \gls{rt} simulations are inherently unlimited in band, i.e. the DS is computed assuming an ideal pulse signal with an unlimited band is transmitted through the multipath channel.
This does not represent a limitation, as the goal of this work is to provide a system-independent characterization of the intrinsic channel properties, without imposing any bandwidth limitations—due to antennas, amplifiers, RF filters, etc.—of any specific communication system.
Nevertheless, the actual channel characteristics of a finite-
bandwidth system can easily be derived, for instance, through the convolution
of the channel’s impulse response with the impulse response of the RF chain. This issue is shortly addressed in the final Appendix.

\subsubsection{Doppler Spread}
\gls{rt} simulations also provide the angle-of-arrival/departure and per-ray power information needed for the computation of the Doppler dispersion of the channel. The Doppler shift experienced by the $i$-th multipath component, accounting for both TX (satellite) and RX motion, is given by \cite{Bilibashi2023DynamicRT}:
\begin{equation}
f_{D,i} = f_0 \left( \frac{c_0 + \mathbf{v}_{\mathrm{RX}} \cdot \hat{\mathbf{k}}_{\mathrm{AoA},i}}{c_0 - \mathbf{v}_{\mathrm{TX}} \cdot \hat{\mathbf{k}}_{\mathrm{AoD},i}} - 1 \right),
\label{eq:doppler_shift}
\end{equation}
where $f_0$ is the carrier frequency, $c_0$ the speed of light in free-space, $\mathbf{v}_{\mathrm{RX}}$ and $\mathbf{v}_{\mathrm{TX}}$ are the RX and TX velocity vectors, and $\hat{\mathbf{k}}_{\mathrm{AoA},i}$, $\hat{\mathbf{k}}_{\mathrm{AoD},i}$ are the unit vectors along the angle-of-arrival and angle-of-departure of the $i$-th ray, respectively. By direct analogy with \eqref{eq:DS}-\eqref{eq:DS_meandelay}, the Doppler Spread $B_D$ is obtained as the power-weighted RMS dispersion of the per-ray Doppler shifts around their mean value $\bar{f}_D$:
\begin{equation}
\mathrm{B}_{D} = \sqrt{\frac{\sum_{i} P_i (f_{D,i} - \bar{f}_D)^2}{\sum_{i} P_i}}, \qquad \bar{f}_D = \frac{\sum_{i} P_i f_{D,i}}{\sum_{i} P_i}.
\label{eq:doppler_spread}
\end{equation}
Unlike the Doppler shift, which simply reflects a frequency translation, Doppler spread quantifies the extent to which this shift varies across different multipath components; it thus represents the parameter that effectively governs the channel's dynamic, time-varying nature, in direct analogy to the role of DS in the delay domain, which is instead related to frequency selectivity.

In this work, the Doppler spread has been estimated assuming the satellite moving along a circular orbit at 500~km altitude (orbital speed of about 7.6~km/s), lying in the vertical plane containing the receiver and the satellite azimuth direction: $\mathbf{v}_{\mathrm{TX}}$ is therefore tangent to the orbit, with horizontal and vertical components depending on the satellite elevation. Two cases are then compared: a static terminal ($|\mathbf{v}_{\mathrm{RX}}| = 0$, i.e. satellite motion only) and a vehicular terminal moving horizontally at $|\mathbf{v}_{\mathrm{RX}}| = 50$~km/h along the street direction, representative of a typical urban vehicular speed.

\section{Results and discussion}
This section includes and discusses some results extracted from \gls{rt} simulations. First, LoS occurrence and the role played by different electromagnetic interactions (reflection, diffraction, and scattering) are investigated for different satellite elevations and propagation scenarios. Then, the addressed propagation parameters (shadowing standard deviation, Ricean \(K\)-factor, delay and Doppler spread) are analyzed at different frequencies and for the considered use-cases, also taking into account their sensitivity to the urban layout.

\subsection{Line of Sight Probability}

\begin{figure}
    \centering
    \includegraphics[width=1\linewidth]{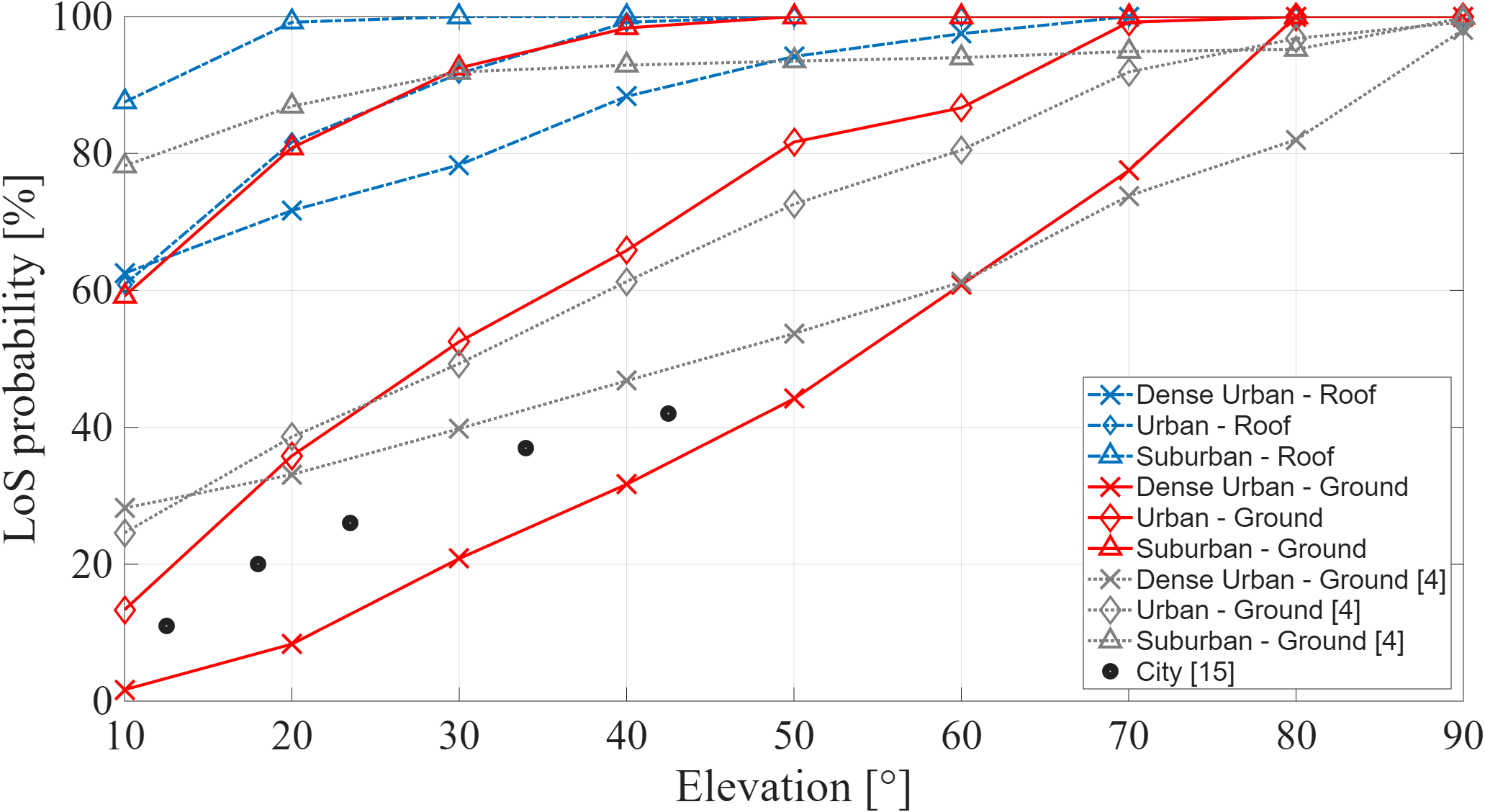}
    \caption{Line of sight probability in different scenarios}
    \label{fig:los_prob}
\end{figure}

LoS probability is easily computed from \gls{rt} simulations, for the different urban environments and deployment of UEs at roof or at street level (Fig. \ref{fig:los_prob}). Regardless of the specific case, LoS occurrence increases with satellite elevation, which is obvious. Moving the ground station from street level to roof height also greatly improves the LoS probability. Moreover, the greater the building density, the lower the LoS probability. For instance, in dense urban environment and user equipment at street level, LoS probability approaches 100\% just when the satellite is close to zenith, whereas full LoS conditions already occur at 40$^{\circ}$ in the suburban case.

It is worth noting that our \gls{rt} simulations are in some disagreement with the LoS probability values suggested in \cite{3gppTR38811}, especially at low satellite elevation in the dense- and sub-urban case. Although no details are provided in \cite{3gppTR38811}, a LoS probability equal to about 25\% in dense urban context when the satellite is at 10$^{\circ}$ elevation appears actually too large. In this regard, on-field evaluations reported in \cite{LandMobileSatCommChannel} (black dots in Fig. \ref{fig:los_prob}) seem in better agreement with \gls{rt} results.

\subsection{Analysis of Propagation Mechanisms}
\begin{figure}[t]
    \centering

    \begin{subfigure}[t]{0.48\linewidth}
        \centering
        \resizebox{\linewidth}{!}{
%
%
\definecolor{mycolor1}{rgb}{0.80000,0.50000,0.00000}%
\definecolor{mycolor2}{rgb}{1.00000,1.00000,0.00000}%
\definecolor{mycolor3}{rgb}{0.00000,1.00000,1.00000}%
\begin{tikzpicture}

\begin{axis}[%
at={(0.758in,0.481in)},
scale only axis,
bar width=0.8,
xmin=0.2,
xmax=4.3,
xtick={1,2,3},
xticklabel style = {font=\large},
xticklabels={{$\text{10}^\circ$},{$\text{50}^\circ$},{$\text{80}^\circ$}},
xlabel style={font=\color{white!15!black}},
xlabel={\Large Satellite Elevation Angle},
ymin=0,
ymax=100,
yticklabel style = {font=\large},
ytick={0,10,20,30,40,50,60,70,80,90,100},
ylabel style={font=\color{white!15!black}},
ylabel={\Large Received Power Percentage (\%)},
axis background/.style={fill=white},
xmajorgrids,
ymajorgrids,
legend style={font=\large,
    at={(0.98,0.02)},
    anchor=south east,
    legend cell align=left,
    draw=white!15!black
}
]
\addplot[ybar stacked, fill=newDeepBlue, draw=black, area legend] table[row sep=crcr] {%
1	1.23692880800406\\
2	36.0990699357781\\
3	79.6678931902734\\
};
\addplot[forget plot, color=white!15!black] table[row sep=crcr] {%
0.2	0\\
4.2	0\\
};
\addlegendentry{L}

\addplot[ybar stacked, fill=pastelred, draw=black, area legend] table[row sep=crcr] {%
1	12.6323974273751\\
2	22.0717663751298\\
3	8.42244507497201\\
};
\addplot[forget plot, color=white!15!black] table[row sep=crcr] {%
0.2	0\\
4.2	0\\
};
\addlegendentry{R}

\addplot[ybar stacked, fill=graphOrange, draw=black, area legend] table[row sep=crcr] {%
1	14.0630235575936\\
2	1.83383645772217\\
3	0.318782144099709\\
};
\addplot[forget plot, color=white!15!black] table[row sep=crcr] {%
0.2	0\\
4.2	0\\
};
\addlegendentry{RD}

\addplot[ybar stacked, fill=corn, draw=black, area legend] table[row sep=crcr] {%
1	21.2670029963731\\
2	23.3960805247013\\
3	4.47828716503046\\
};
\addplot[forget plot, color=white!15!black] table[row sep=crcr] {%
0.2	0\\
4.2	0\\
};
\addlegendentry{D}

\addplot[ybar stacked, fill=emeraldGreen, draw=black, area legend] table[row sep=crcr] {%
1	30.4836493584873\\
2	12.3218686572241\\
3	6.74454625599954\\
};
\addplot[forget plot, color=white!15!black] table[row sep=crcr] {%
0.2	0\\
4.2	0\\
};
\addlegendentry{S}

\addplot[ybar stacked, fill=lightskyblue, draw=black, area legend] table[row sep=crcr] {%
1	20.3169978521669\\
2	4.27737804944446\\
3	0.368046169624929\\
};
\addplot[forget plot, color=white!15!black] table[row sep=crcr] {%
0.2	0\\
4.2	0\\
};
\addlegendentry{RS}

\end{axis}
\end{tikzpicture}%
}
        \caption{Dense Urban – 3.5 GHz}
        \label{fig:denseurban_3_5}
    \end{subfigure}
    \hfill
    \begin{subfigure}[t]{0.48\linewidth}
        \centering
        \resizebox{\linewidth}{!}{
%
%
\definecolor{mycolor1}{rgb}{0.80000,0.50000,0.00000}%
\definecolor{mycolor2}{rgb}{1.00000,1.00000,0.00000}%
\definecolor{mycolor3}{rgb}{0.00000,1.00000,1.00000}%
\begin{tikzpicture}

\begin{axis}[%
at={(0.758in,0.481in)},
scale only axis,
bar width=0.8,
xmin=0.2,
xmax=4.3,
xtick={1,2,3},
xticklabel style = {font=\large},
xticklabels={{$\text{10}^\circ$},{$\text{50}^\circ$},{$\text{80}^\circ$}},
xlabel style={font=\color{white!15!black}},
xlabel={\Large Satellite Elevation Angle},
ymin=0,
ymax=100,
yticklabel style = {font=\large},
ytick={0,10,20,30,40,50,60,70,80,90,100},
ylabel style={font=\color{white!15!black}},
ylabel={\Large Received Power Percentage (\%)},
axis background/.style={fill=white},
xmajorgrids,
ymajorgrids,
legend style={font=\large,
    at={(0.98,0.02)},
    anchor=south east,
    legend cell align=left,
    draw=white!15!black
}
]
\addplot[ybar stacked, fill=newDeepBlue, draw=black, area legend] table[row sep=crcr] {%
1	1.30886165721046\\
2	35.6853291362738\\
3	77.1724944145437\\
};
\addplot[forget plot, color=white!15!black] table[row sep=crcr] {%
0.2	0\\
4.2	0\\
};
\addlegendentry{L}

\addplot[ybar stacked, fill=pastelred, draw=black, area legend] table[row sep=crcr] {%
1	11.492362601128\\
2	18.2206145158427\\
3	5.28690110193151\\
};
\addplot[forget plot, color=white!15!black] table[row sep=crcr] {%
0.2	0\\
4.2	0\\
};
\addlegendentry{R}

\addplot[ybar stacked, fill=graphOrange, draw=black, area legend] table[row sep=crcr] {%
1	2.47548373346558\\
2	0.60077393975724\\
3	0.062680387374049\\
};
\addplot[forget plot, color=white!15!black] table[row sep=crcr] {%
0.2	0\\
4.2	0\\
};
\addlegendentry{RD}

\addplot[ybar stacked, fill=corn, draw=black, area legend] table[row sep=crcr] {%
1	7.30603671660285\\
2	9.73614333119789\\
3	1.47073461493615\\
};
\addplot[forget plot, color=white!15!black] table[row sep=crcr] {%
0.2	0\\
4.2	0\\
};
\addlegendentry{D}

\addplot[ybar stacked, fill=emeraldGreen, draw=black, area legend] table[row sep=crcr] {%
1	50.9568080271768\\
2	28.543230239371\\
3	15.4588118958084\\
};
\addplot[forget plot, color=white!15!black] table[row sep=crcr] {%
0.2	0\\
4.2	0\\
};
\addlegendentry{S}

\addplot[ybar stacked, fill=lightskyblue, draw=black, area legend] table[row sep=crcr] {%
1	26.4604472644163\\
2	7.21390883755742\\
3	0.548377585406257\\
};
\addplot[forget plot, color=white!15!black] table[row sep=crcr] {%
0.2	0\\
4.2	0\\
};
\addlegendentry{RS}

\end{axis}
\end{tikzpicture}%
}
        \caption{Dense Urban – 40 GHz}
        \label{fig:denseurban_40}
    \end{subfigure}

    \vspace{0.5em}

    \begin{subfigure}[t]{0.48\linewidth}
        \centering
        \resizebox{\linewidth}{!}{
%
%
\definecolor{mycolor1}{rgb}{0.80000,0.50000,0.00000}%
\definecolor{mycolor2}{rgb}{1.00000,1.00000,0.00000}%
\definecolor{mycolor3}{rgb}{0.00000,1.00000,1.00000}
\begin{tikzpicture}

\begin{axis}[%
at={(0.758in,0.481in)},
scale only axis,
bar width=0.8,
xmin=0.2,
xmax=4.3,
xticklabel style = {font=\large},
xtick={1,2,3},
xticklabels={{$\text{10}^\circ$},{$\text{50}^\circ$},{$\text{80}^\circ$}},
xlabel style={font=\color{white!15!black}},
xlabel={\Large Satellite Elevation Angle},
ymin=0,
ymax=100,
yticklabel style = {font=\large},
ytick={0,10,20,30,40,50,60,70,80,90,100},
ylabel style={font=\color{white!15!black}},
ylabel={\Large Received Power Percentage (\%)},
axis background/.style={fill=white},
xmajorgrids,
ymajorgrids,
legend style={font=\large,
    at={(0.98,0.02)},
    anchor=south east,
    legend cell align=left,
    draw=white!15!black
}
]
\addplot[ybar stacked, fill=newDeepBlue, draw=black, area legend] table[row sep=crcr] {%
1	46.5289341059617\\
2	90.2197997096809\\
3	85.7981661545521\\
};
\addplot[forget plot, color=white!15!black] table[row sep=crcr] {%
0.2	0\\
4.2	0\\
};
\addlegendentry{L}

\addplot[ybar stacked, fill=pastelred, draw=black, area legend] table[row sep=crcr] {%
1	21.289894357254\\
2	6.77173566538935\\
3	11.4164272559552\\
};
\addplot[forget plot, color=pastelred] table[row sep=crcr] {%
0.2	0\\
4.2	0\\
};
\addlegendentry{R}

\addplot[ybar stacked, fill=graphOrange, draw=black, area legend] table[row sep=crcr] {%
1	3.01768122387483\\
2	0.0969411064619155\\
3	0.0271172187757715\\
};
\addplot[forget plot, color=graphOrange] table[row sep=crcr] {%
0.2	0\\
4.2	0\\
};
\addlegendentry{RD}

\addplot[ybar stacked, fill=corn, draw=black, area legend] table[row sep=crcr] {%
1	23.3263879100081\\
2	0.66590727335698\\
3	0.198866099800279\\
};
\addplot[forget plot, color=corn] table[row sep=crcr] {%
0.2	0\\
4.2	0\\
};
\addlegendentry{D}

\addplot[ybar stacked, fill=emeraldGreen, draw=black, area legend] table[row sep=crcr] {%
1	4.80973548189393\\
2	2.14385136484774\\
3	2.46630920043341\\
};
\addplot[forget plot, color=emeraldGreen!15!black] table[row sep=crcr] {%
0.2	0\\
4.2	0\\
};
\addlegendentry{S}

\addplot[ybar stacked, fill=lightskyblue, draw=black, area legend] table[row sep=crcr] {%
1	1.02736692100744\\
2	0.101764880263087\\
3	0.0931140704832769\\
};
\addplot[forget plot, color=lightskyblue] table[row sep=crcr] {%
0.2	0\\
4.2	0\\
};
\addlegendentry{RS}

\end{axis}

\end{tikzpicture}%
}
        \caption{Suburban – 3.5 GHz}
        \label{fig:suburban_3_5}
    \end{subfigure}
    \hfill
    \begin{subfigure}[t]{0.48\linewidth}
        \centering
        \resizebox{\linewidth}{!}{
%
%
\definecolor{mycolor1}{rgb}{0.80000,0.50000,0.00000}%
\definecolor{mycolor2}{rgb}{1.00000,1.00000,0.00000}%
\definecolor{mycolor3}{rgb}{0.00000,1.00000,1.00000}%
\begin{tikzpicture}

\begin{axis}[%
at={(0.758in,0.481in)},
scale only axis,
bar width=0.8,
xmin=0.2,
xmax=4.3,
xticklabel style = {font=\large},
xtick={1,2,3},
xticklabels={{$\text{10}^\circ$},{$\text{50}^\circ$},{$\text{80}^\circ$}},
xlabel style={font=\color{white!15!black}},
xlabel={\Large Satellite Elevation Angle},
ymin=0,
ymax=100,
yticklabel style = {font=\large},
ytick={0,10,20,30,40,50,60,70,80,90,100},
ylabel style={font=\color{white!15!black}},
ylabel={\Large Received Power Percentage (\%)},
axis background/.style={fill=white},
xmajorgrids,
ymajorgrids,
legend style={font=\large,
    at={(0.98,0.02)},
    anchor=south east,
    legend cell align=left,
    draw=white!15!black
}
]
\addplot[ybar stacked, fill=newDeepBlue, draw=black, area legend] table[row sep=crcr] {%
1	48.1004017221099\\
2	87.5554864605271\\
3	82.6633874399415\\
};
\addplot[forget plot, color=white!15!black] table[row sep=crcr] {%
0.2	0\\
4.2	0\\
};
\addlegendentry{L}

\addplot[ybar stacked, fill=pastelred, draw=black, area legend] table[row sep=crcr] {%
1	23.003309377686\\
2	6.91028354550086\\
3	11.1995291999924\\
};
\addplot[forget plot, color=white!15!black] table[row sep=crcr] {%
0.2	0\\
4.2	0\\
};
\addlegendentry{R}

\addplot[ybar stacked, fill=graphOrange, draw=black, area legend] table[row sep=crcr] {%
1	1.70224624058499\\
2	0.0507211154509806\\
3	0.00573321546376748\\
};
\addplot[forget plot, color=white!15!black] table[row sep=crcr] {%
0.2	0\\
4.2	0\\
};
\addlegendentry{RD}

\addplot[ybar stacked, fill=corn, draw=black, area legend] table[row sep=crcr] {%
1	9.93585158711913\\
2	0.0996126736629183\\
3	0.054084900170229\\
};
\addplot[forget plot, color=white!15!black] table[row sep=crcr] {%
0.2	0\\
4.2	0\\
};
\addlegendentry{D}

\addplot[ybar stacked, fill=emeraldGreen, draw=black, area legend] table[row sep=crcr] {%
1	14.3862454349108\\
2	5.15269268766343\\
3	5.86853786250434\\
};
\addplot[forget plot, color=white!15!black] table[row sep=crcr] {%
0.2	0\\
4.2	0\\
};
\addlegendentry{S}

\addplot[ybar stacked, fill=lightskyblue, draw=black, area legend] table[row sep=crcr] {%
1	2.87194563758915\\
2	0.231203517194667\\
3	0.208727381927711\\
};
\addplot[forget plot, color=white!15!black] table[row sep=crcr] {%
0.2	0\\
4.2	0\\
};
\addlegendentry{RS}

\end{axis}

\end{tikzpicture}%
}
        \caption{Suburban – 40 GHz}
        \label{fig:suburban_40}
    \end{subfigure}

    \caption{Stacked power contribution by propagation mechanisms at satellite elevation angles 10°, 50°, and 80° for Dense Urban and Suburban scenarios at 3.5 GHz and 40 GHz. L – LoS, R – Reflections, RD – Reflections combined with Diffractions, D – Diffractions, S – Diffuse Scattering, RS – Reflections combined with Diffuse Scattering.}
    \label{fig:stacked_mechanisms_halfcol}
\end{figure}

Fig.~\ref{fig:stacked_mechanisms_halfcol} presents stacked bar charts illustrating the average percentage contribution to the total received power of different propagation mechanisms. The analysis includes two reference scenarios, dense urban and suburban, evaluated at two frequencies (3.5 GHz and 40 GHz) and three satellite elevation angles (10°, 50°, and 80°). Circular polarization is assumed for the satellite, while vertical polarization is used for the receivers, located at ground level.

\subsubsection{Dense Urban Scenario}

In the dense urban scenario at 3.5 GHz (Fig.~\ref{fig:denseurban_3_5}), the received signal at low satellite elevation angles (e.g., 10°) is characterized by an almost complete absence of the LoS component (Fig.~\ref{fig:los_prob}). In such conditions, the dominant contributions come from diffuse scattering and scattering combined with reflections. As explained in \cite{DegliEsposti2004}, pure reflections, while possible, require multiple bounces to reach the receiver due to the geometry and height of buildings, leading to high path loss, as for diffractions the receivers may be often located in a deep shadow region.
As the elevation angle increases to 50° and 80°, the LoS probability increases as well,  becoming dominant and contributing to approximately 80\% of the total received power. Reflections contribute to around 10\%, diffractions about 3\%, and the remaining 7\% comes from scattering-related mechanisms.

At 40 GHz (Fig.~\ref{fig:denseurban_40}), the main difference in the dense urban scenario is an increased contribution from diffuse scattering for all elevation angles at the expense of the other mechanisms, especially diffraction, as shorter wavelengths generate less diffraction and more surface-roughness scattering.

\subsubsection{Suburban Scenario}

In the suburban scenario, the propagation environment is characterized by a very different profile. At 3.5 GHz (Fig.~\ref{fig:suburban_3_5}) and 10° elevation, the LoS component already contributes nearly 50\% of the received power. The remaining power is mainly attributed to reflections and diffractions. The diffuse scattering component is less pronounced compared to the dense urban scenario.
As the elevation angle increases, the LoS component quickly becomes dominant, reaching 90\% of the received power at 50°, while the contribution from other mechanisms becomes minor. At 80° elevation, the reflection component gains a few percentage points relative to the LoS, due to the increase in intensity of ground reflections.

At 40 GHz (Fig.~\ref{fig:suburban_40}), similar trends are observed, although scattering plays a more prominent role due to the already discussed surface roughness effects. At 10°, LoS contributes about 45\%, and this rises above 80\% at higher elevation angles. Reflections and diffractions provide only minor contributions, showing that although frequency affects the mechanism balance, the scenario itself remains the most influential factor.

As the UEs are raised at (average) rooftop level, \gls{rt} simulations show that the percentage of power carried by the direct path gets dominant in every scenario and basically at any elevation angle, as the LoS probability is always greater than 60\% (Fig. \ref{fig:los_prob}). The general prevalence of the LoS component is further stressed by the use of highly directive radiation patterns (Table \ref{tab:use_cases}). No figures have been included to investigate the high-rise EU case, since there is basically no competition between the interaction mechanisms to be highlighted.

\subsection{Analysis of Propagation Parameters}

The major results achieved for the target propagation parameters are reported herein. In particular, Rician \(K\)-factor and delay spread are discussed together, as they turned out to be closely related to each other.

\subsubsection{Shadowing}
\begin{figure}
    \centering
    \includegraphics[width=0.8\linewidth]{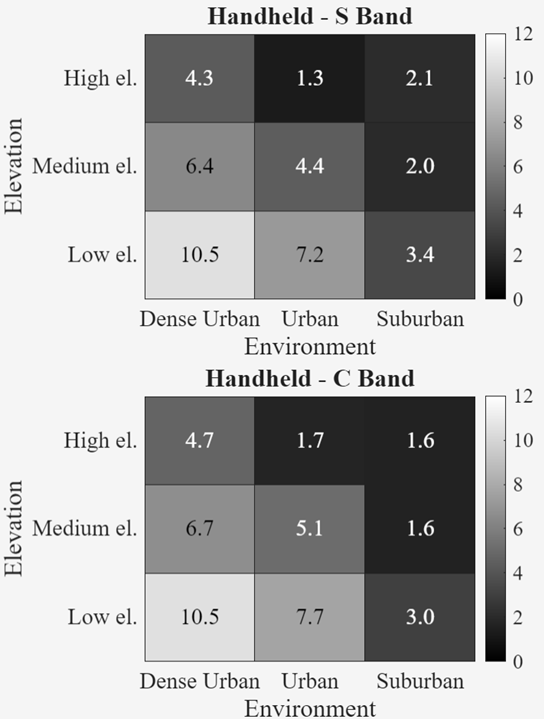}
    \caption{Shadowing std. dev. at different elevations and in different urban contexts for the handheld use-case.}
    \label{fig:HMAP_HH}
\end{figure}

\begin{figure}
    \centering
    \includegraphics[width=0.8\linewidth]{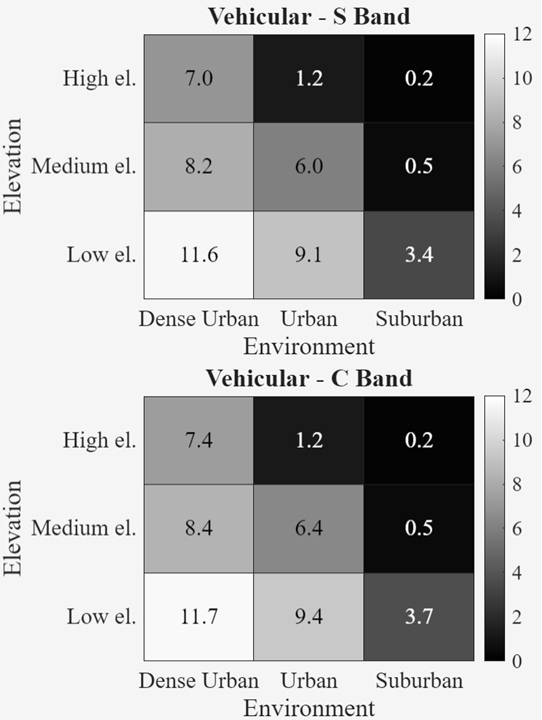}
    \caption{Shadowing std. dev. at different elevations and in different urban contexts for the vehicular use-case.}
    \label{fig:HMap_Veh}
\end{figure}

the values of $\sigma_{dB}$ are reported in Fig. \ref{fig:HMAP_HH} for the handheld case at different elevation levels and for the different urban scenarios. Not surprisingly, a lower LoS probability (lower satellite elevation and/or greater building density) yields heavier shadowing effects, as it entails a greater variability in the propagation conditions at the different ground grids. By contrast, when LoS occurs more frequently, propagation approaches the free space case, with fewer and weaker expected deviations from it. Overall, the difference between the S and the C band is quite small and remains below 1 dB in all considered cases. The Vehicular use case is investigated in Fig. \ref{fig:HMap_Veh}, showing the same sensitivity to urban density and elevation.

It is worth noting that under prevailing LoS conditions all over the grids (i.e., LoS probability $\approx$ 100\%), an increased antenna directivity results in lower shadowing effect, provided that, of course, the directive antenna is effectively pointed in the direction of the link. In that case, in fact, the received power becomes increasingly dominated by a component (the LoS contribution)  that is relatively stable across the grids, and deviations from this general condition are uncommon. This is confirmed by the comparison between the handheld and the vehicular case, for the same band, with reference to the urban case at high elevation, and to the suburban case for medium and high elevation, where the LoS probability is actually $\approx$ 100\% (Fig. \ref{fig:los_prob}). By contrast, when the occurrence of NLoS conditions is not negligible, an antenna with greater directivity but ineffectively pointed in the NLoS cases, corresponds to increased shadowing effects. NLoS reception with a mis-aligned antenna relies on indirect paths, and the received power becomes more sensitive to whether the available indirect paths arrive from directions with high or low antenna gain. A greater variability exists among the local mean powers at the different grids, thus resulting in a greater $\sigma_{dB}$ value. This actually results in the comparison between Fig. \ref{fig:HMAP_HH} and Fig. \ref{fig:HMap_Veh} in every case where the NLoS probability is not negligible.
This trend related to the directivity of the ground antenna becomes even more pronounced when the Fixed use-case is considered. Since the antenna of the ground station is then extremely directive, the value of $\sigma_{dB}$ drops to almost zero under dominant LoS conditions, whereas it grows up to awkwardly large values (tens of dB) otherwise. Since this represents a limit case, the corresponding table has not been included.

\subsubsection{Rician \(K\)-factor and Delay Spread} 
\begin{figure}
    \centering
    \includegraphics[width=1\linewidth]{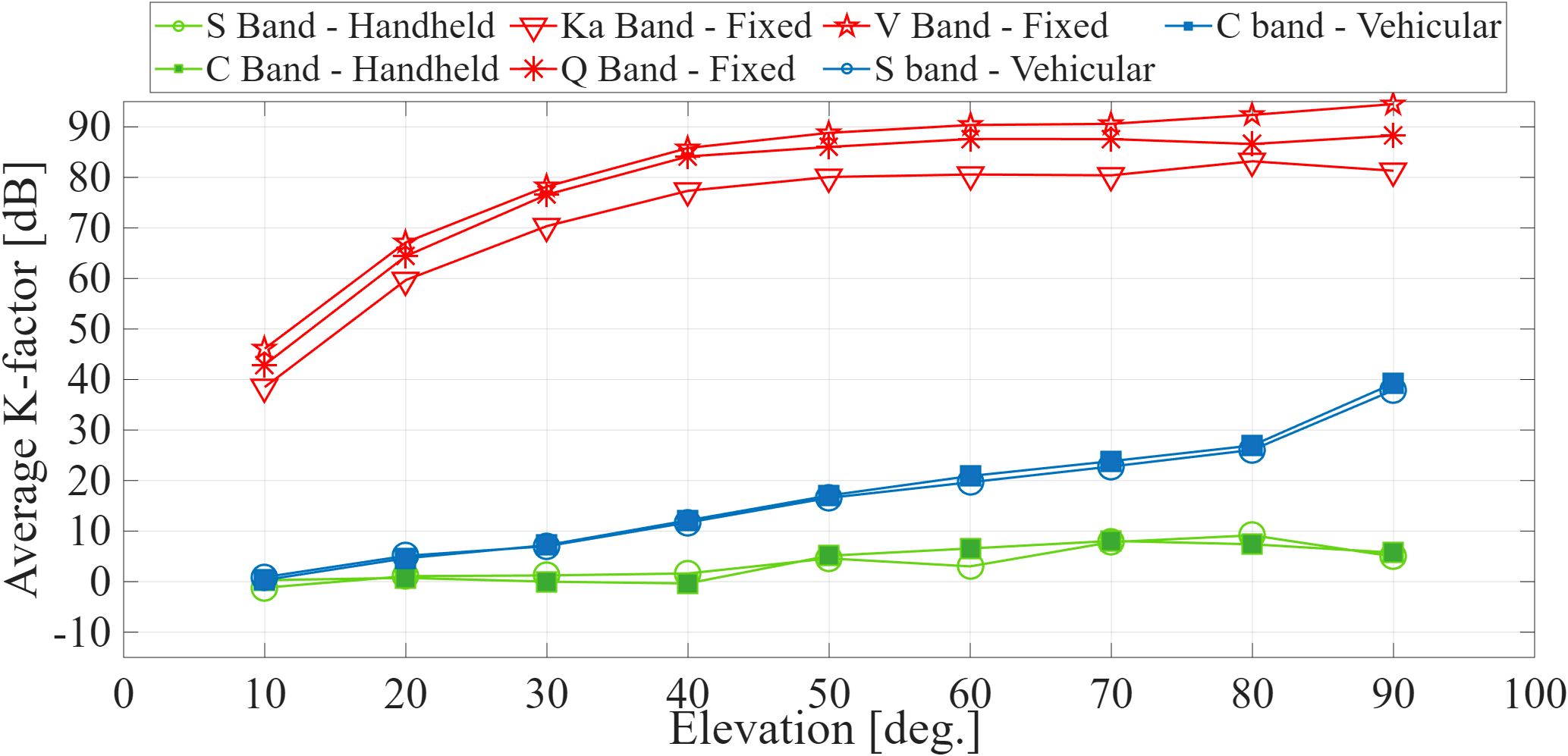}
    \caption{Rician \(K\)-factor vs. satellite elevation, urban environment}
    \label{fig:K_vs_el}
\end{figure}

\begin{figure}
    \centering
    \includegraphics[width=1\linewidth]{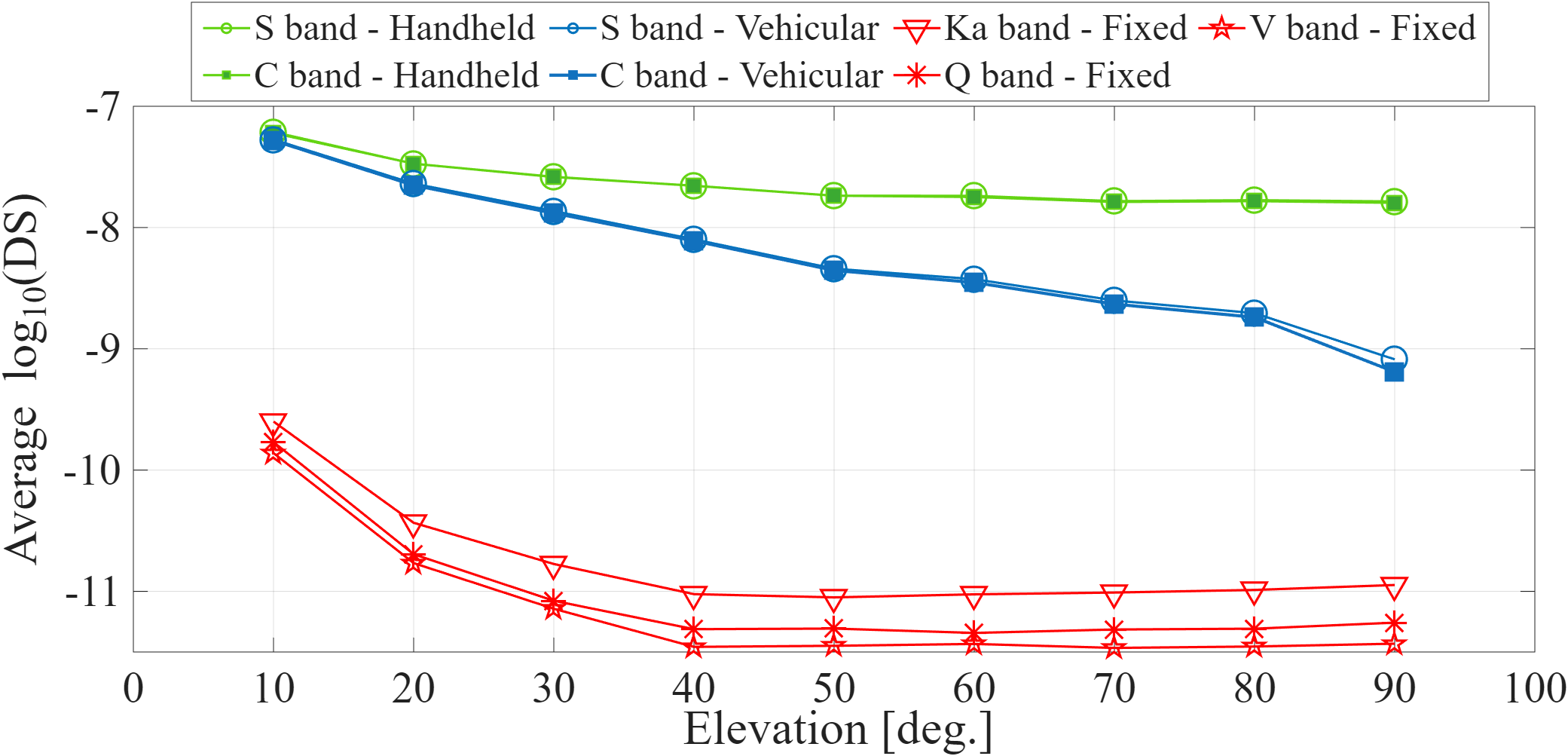}
    \caption{Delay Spread vs. satellite elevation, urban environment}
    \label{fig:DS_vs_el}
\end{figure}

they are reported in Figs. \ref{fig:K_vs_el}, \ref{fig:DS_vs_el} against the satellite elevation in urban environment, in the considered frequency bands, and for the different use cases. Similar trends have been obtained for the dense- and the sub-urban scenarios.

Both Rician \(K\)-factor and delay spread are in general clearly affected by the occurrence of the LoS condition. This makes physical sense, as greater LoS probability entails the presence of a powerful, minimum-delay received signal contribution with higher likelihood. At the same time, it does not automatically trigger a corresponding, equivalent increase in the power carried by scattered multipath components arriving with long(er) delay. In fact, multipath structure and richness are greatly affected by the density of items and obstructions, which is not expected to change a lot within the same environment. These considerations provide a general explanation for the reason why a greater LoS probability contributes – at least on the average – to increase the Rician \(K\)-factor and to reduce the delay spread. In the framework of the considered \gls{rt} simulations, two main factors have a clear impact on the LoS condition (and therefore on the values of \(K\)-factor and DS):
\begin{itemize}
    \item \textit{The elevation angle}, as of course the satellite achieves easier visibility of the ground stations when it flies at higher elevation. Close to zenith (80$^{\circ}$-90$^{\circ}$) the Rician \(K\)-factor can be 10 to 30 dB greater than its value at the lowest elevation (Fig. \ref{fig:K_vs_el}), whereas the delay spread can decrease by one to two orders of magnitude (Fig. \ref{fig:DS_vs_el});
    \item \textit{The height of the earth station}: if moved from street- to roof-level, the Rician \(K\)-factor increases by more than 40dB (Fig. \ref{fig:K_vs_el}), whereas the delay spread reduces by two to three orders of magnitude (Fig. \ref{fig:DS_vs_el}). Although the significant difference in \(K\)-factor and DS values between the handheld/vehicular case (antenna at ground level) and the fixed case (antenna at rooftop level) is also due to the different radiation pattern (Table \ref{tab:use_cases}), the difference in antenna height is expected to represent a crucial aspect;
\end{itemize}

To some extent, the role played by the antenna radiation pattern alone stands out in the comparison between the handheld and the vehicular case, where the antenna is always at ground level. A greater directivity makes a difference (up to about 10-15 dB for \(K\)-factor and up to one order of magnitude for DS) at larger elevation angles (40$^{\circ}$-90$^{\circ})$, where the vertical boresight direction of the patch antenna more easily corresponds to the LoS, dominant path, therefore boosting its intensity at the expense of the other scattered paths with longer propagation delay and direction of arrival outside the main radiation lobe.

With reference to the frequency sensitivity, not surprisingly S and C bands basically yield the same values (Figs. \ref{fig:DS_vs_el},\ref{fig:K_vs_el}), i.e. they share the same propagation conditions. The same remark holds for the Q and the V bands, though to a lesser extent. In comparison with the Ka band, the Rician \(K\)-factor at Q/V frequencies is about 5 dB greater, whereas only a slight difference is observed in the delay spread.

\begin{figure}
    \centering
    \includegraphics[width=1\linewidth]{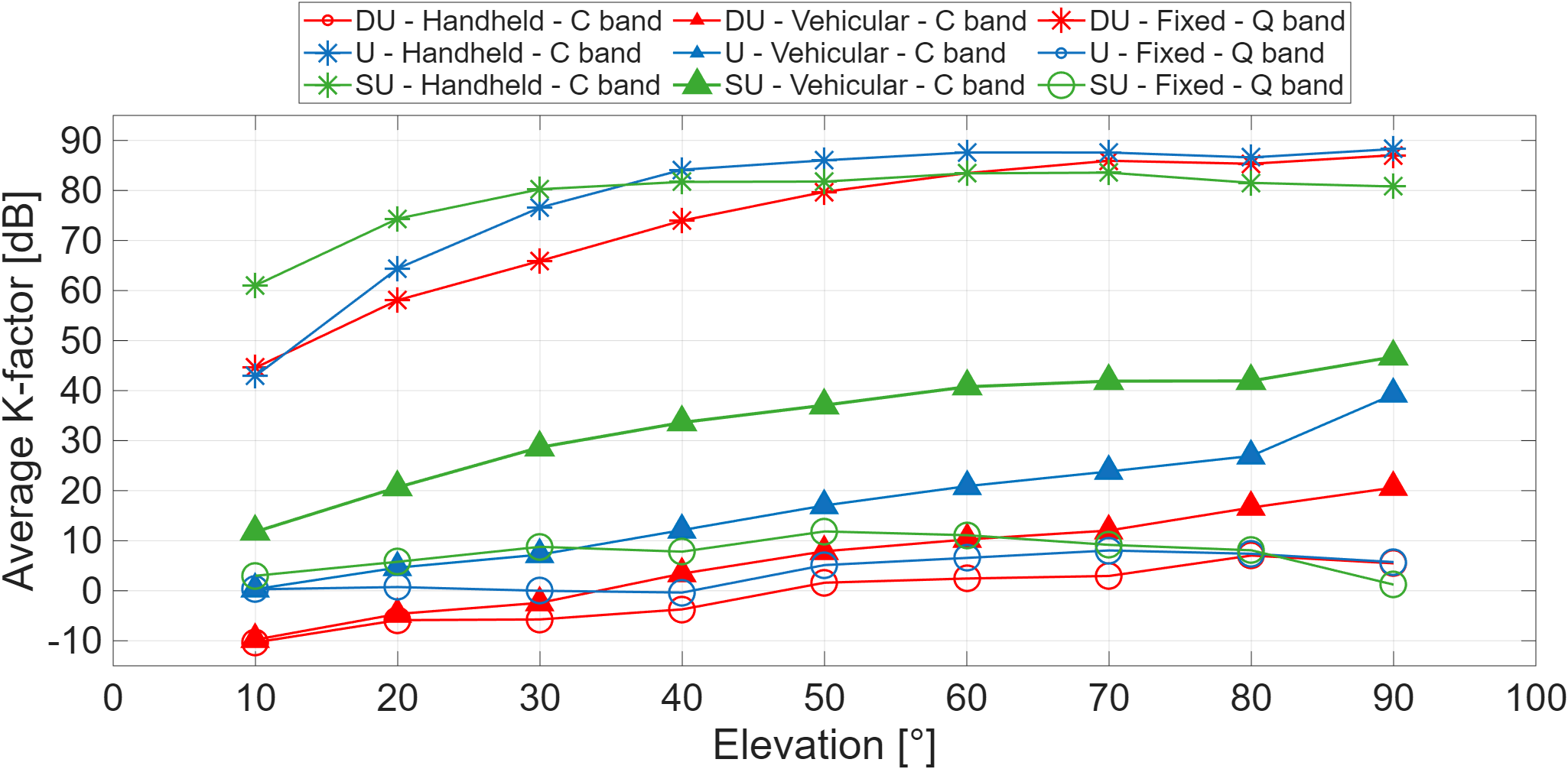}
    \caption{Rician \(K\)-factor vs. satellite elevation, sensitivity to different environments}
    \label{fig:K_vs_env}
\end{figure}

The impact of the urban context type on the considered propagation parameters is analyzed in Figs. \ref{fig:K_vs_env}, \ref{fig:DS_vs_env}. Generally speaking, the curves related to the different urban scenarios (dense urban, urban and suburban) for the same frequency and use case are often somehow interleaved, meaning that the mechanisms that generate multipath richness and dispersion, and therefore affect \(K\)-factor and DS values, appear to be quite complex.
This is not a relevant issue when the UEs are located at building levels (lines marked with stars in Figs. \ref{fig:K_vs_env}, \ref{fig:DS_vs_env}), as the Rician \(K\)-factor values are then so large (several tens of dB) and the corresponding DS values are so small ($< 10^{-10}$ sec. on average) that time dispersion and fast fading look like really minor effects. Wireless propagation occurs as in free space conditions to a great extent.
By contrast, when the UEs height is moved to street level, the free space approximation no longer holds and some interpretation of results might be of interest. In this respect, the existence of dominant, minimum-delay rays -- such as the direct ray and single-bounce rays reflected by vertical building walls, which tend to suppress multipath richness, appears to be a discriminating factor at low satellite elevation angles. In this case, increasing the elevation from 10° to 30–40° raises the LoS probability and therefore decreases multipath richness, with a corresponding increase in the Rician \(K\)-factor and decrease in the DS. Moreover, suburban and urban environments appear to exhibit lower multipath richness than dense urban environments, where tall buildings often obstruct dominant rays -- as expected.

The situation is different at high elevation angles. In this case, the LoS probability is very high across all urban layouts, and the degree of multipath richness, although low, is mainly determined by ground-level scattering due to clutter and vehicles. In the extreme case when elevation is close to the zenith, vertical walls are not "seen" from the satellite, and therefore they don't generate any multipath. Under these conditions, suburban environments often show a lower Rician \(K\)-factor and higher delay spread than dense urban layouts, likely because the lower buildings density allows scattering from more distant terrain spots to be visible, thereby increasing richness and dispersion. This may result in a crossover between the suburban curves (green lines) and the other curves in Figs. \ref{fig:K_vs_env} and \ref{fig:DS_vs_env}, due to a milder slope factor in the suburban case. Parameter variation with elevation, however, is not very pronounced at high elevation angles.

A negligible sensitivity of both Rician \(K\)-factor and DS to the propagation scenario can be also noted for the handheld case (lines marked with empty circles in Figs. \ref{fig:K_vs_env} and \ref{fig:DS_vs_env}), where an isotropic radiation pattern is considered at the UE. A possible explanation is again related to the LoS probability, that of course increases, irrespective of the elevation, moving from dense urban to urban and suburban layouts due to the lower building density. Generally speaking, a greater LoS occurrence across different environments is expected to produce a twofold effect. On the one hand, a more frequent existence of a dominant propagation path that is weakly obstructed or entirely unobstructed; on the other hand, a stronger multipath richness, as a sparser building layout opens the way to a larger number of multipath contributions with still significant intensity in spite of the long(er) propagation delays. The former effect of course contributes to increase the Rician \(K\)-factor and reduce the delay spread, whereas the latter works in the opposite direction. According to the \gls{rt} simulations, a sort of balance between these two effects takes place, resulting in very close \(K\)-factor and DS values across the different urban layouts.

It is worth pointing out that this result is somehow in contrast with the trend reported in \cite{3gppTR38811}, which highlights a difference of about one order of magnitude in delay spread between the dense urban and the suburban case up to 60 deg. elevation. Providing a clear explanation for such difference is hardly possible, as delay spread values in \cite{3gppTR38811} are reported as tabular data without any detail about their computation procedure. As far as the Rician \(K\)-factor is concerned, a fair comparison is not possible, as its assessment is limited in \cite{3gppTR38811} to LoS conditions only, although LoS occurrence does not necessarily represent the standard case when the UE is located at street level. Anyway, some counter-intuitive trends for the Rician \(K\)-factor values reported in \cite{3gppTR38811} appear evident, as already highlighted in \cite{EuCAP2024_satellite}.


The use of a receiving antenna with greater directivity and vertical boresight (vehicular case in Figs. \ref{fig:K_vs_env} and \ref{fig:DS_vs_env}) reduces the lateral scattering contributions reaching the UEs through the gaps in the building layer, while preserving the intensity of dominant contributions arriving from higher elevation angles. Therefore, Rician \(K\)-factor values in the dense urban scenario remain smaller than those in the suburban environment, with the urban case in the middle (lines marked with triangles in Figs. \ref{fig:K_vs_env} and \ref{fig:DS_vs_env}). A corresponding reversed trend applies to DS values.

\subsubsection{Doppler Spread}
The previous parameters are related to large- and small-scale fading characteristics and to the average, static channel dispersion, but do not directly address how quickly the channel changes over time.
In this respect, Table~\ref{tab:doppler_spread_env} reports the resulting Doppler Spread, averaged over the 20 receiver grids, at 3.5 GHz, for the three considered urban contexts and three representative elevation angles. Moreover, the static-terminal case ($|\mathbf{v}_{\mathrm{RX}}|$ = 0, i.e. satellite motion only) and the vehicular ground terminal ($|\mathbf{v}_{\mathrm{RX}}|$ = 50 km/h), are also compared in the Table.

\begin{table}[t]
\centering
\caption{Doppler spread (mean over 20 grid positions) at 3.5~GHz, static case ($v_{\mathrm{RX}} = 0$~km/h) vs. urban vehicular case ($v_{\mathrm{RX}} = 50$~km/h), across three environment types and three elevation angles. All values in Hz.}
\label{tab:doppler_spread_env}
\begin{tabular}{l cc cc cc}
\toprule
& \multicolumn{2}{c}{$20^\circ$} & \multicolumn{2}{c}{$50^\circ$} & \multicolumn{2}{c}{$80^\circ$} \\
\cmidrule(lr){2-3}\cmidrule(lr){4-5}\cmidrule(lr){6-7}
Environment & Static & Vehic. & Static & Vehic. & Static & Vehic. \\
\midrule
Dense Urban & $2.5$ & $56.4$ & $1.0$ & $20.4$ & $0.4$ & $7.6$ \\
Urban  & $1.3$ & $44.3$ & $0.7$ & $12.2$ & $0.6$ & $3.6$ \\
Suburban & $1.3$ & $28.5$ & $0.7$ & $2.2$ & $0.7$ & $1.1$ \\
\bottomrule
\end{tabular}
\end{table}

Despite the considerable absolute speed of the satellite, the static-case Doppler Spread in Table~\ref{tab:doppler_spread_env} is only a few Hz at most, and falls below 1 Hz at the higher elevations. This confirms that satellite motion alone, in spite of the large Doppler shift it produces, translates into only a modest dispersion of that shift across the multipath components -- and therefore into a weakly time-varying channel -- because at a distance of 500 km the angular spread of the departure directions subtended by the urban environment is small, so that all rays experience an almost identical projected satellite velocity. The trend with elevation mirrors that already observed for the $K$-factor and DS (Figs.~\ref{fig:K_vs_el}, \ref{fig:DS_vs_el}): as elevation increases, the LoS condition becomes dominant and multipath richness decreases, further narrowing the Doppler-shift dispersion. Furthermore, the Doppler Spread generally tends to increase with the complexity of the environment, i.e. moving from Suburban to Urban and then to the Dense Urban case. A marginal reversal of the environment ranking can also be observed at the highest elevation (0.7 Hz for Suburban vs. 0.4 Hz for Dense Urban), though at such sub-Hz levels the difference is of little practical relevance.

\begin{figure}
    \centering
    \includegraphics[width=1\linewidth]{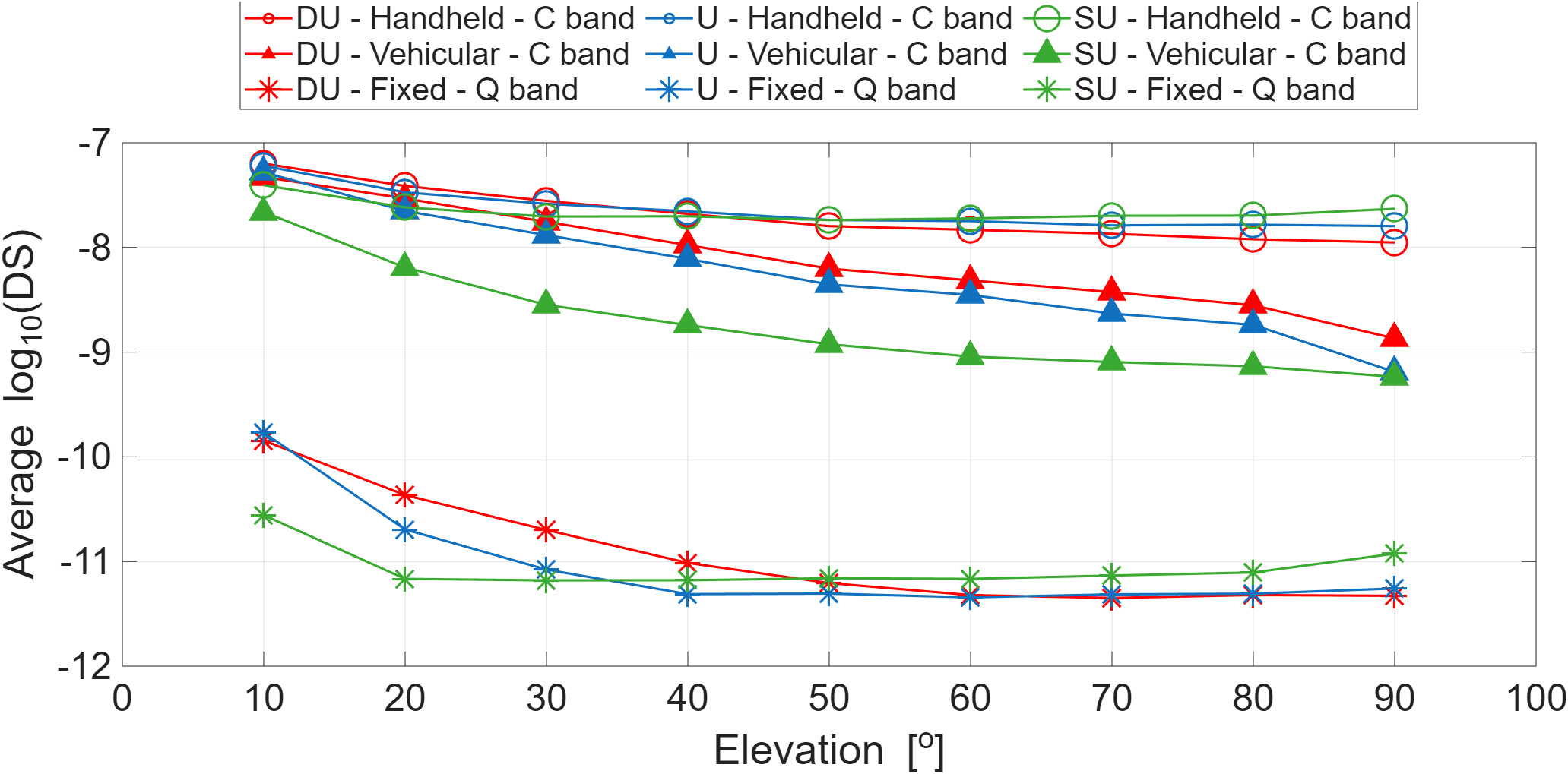}
    \caption{Delay Spread vs. satellite elevation, sensitivity to different environment}
    \label{fig:DS_vs_env}
\end{figure}

When the RX terminal itself is in motion, the Doppler Spread increases considerably with respect to the static case, since a moderate terminal speed introduces, through the $\mathbf{v}_{\mathrm{RX}} \cdot \hat{\mathbf{k}}_{\mathrm{AoA},i}$ term in \eqref{eq:doppler_shift}, a Doppler contribution that varies markedly with the angle of arrival of each multipath component -- i.e., with the very same multipath richness that governs the delay spread. Consistently with this interpretation, the vehicular Doppler Spread decreases with elevation and is consistently larger in the dense urban scenario than in the urban and suburban ones, following the same environment ranking discussed above for the DS (Fig.~\ref{fig:DS_vs_env}). Even so, the vehicular Doppler Spread remains modest in absolute terms, not exceeding about 56 Hz even in the worst case (dense urban, low elevation); at 3.5 GHz, this corresponds to coherence times on the order of tens of milliseconds, well above typical physical-layer processing intervals. Overall, these results indicate that the dynamicity of the simulated satellite-to-urban channel is chiefly governed by terminal mobility rather than by the (much faster, but far less angularly dispersive) satellite motion, and that, even for a vehicular terminal, the channel remains time-invariant over intervals long enough for the average $K$-factor/DS characterization above to retain practical relevance.

\section{Conclusions}
This work investigates satellite-to-urban propagation through deterministic ray-tracing simulations, aiming to determine dominant propagation mechanisms and characterize major propagation parameters (shadowing standard deviation,
Rician \(K\)-factor, RMS delay spread and Doppler spread) under different urban layouts, user equipment configurations, and operating frequencies. The analysis highlights the strong dependence of these channel parameters on LoS/NLoS conditions, which are primarily governed by satellite elevation and receiver placement.

Results show that increasing the satellite elevation significantly improves LoS probability, leading to weaker shadowing effects, larger Rician \(K\)-factor and reduced delay and Doppler spread, as expected. The position of the user equipment also plays a major role: moving the receiver from street level to rooftop height substantially increases the Rician \(K\)-factor and strongly decreases both shadowing and delay spread due to the higher likelihood of unobstructed propagation.

Antenna characteristics further influence the channel statistics. Directive antennas with vertical boresight enhance the dominant path contribution while suppressing lateral multipath components.
This results in a reduced shadowing standard deviation only under predominantly LoS conditions, while the impact is instead reversed otherwise.
Greater antenna directivity also corresponds to higher Rician \(K\)-factor and smaller delay spread compared to isotropic reception.

While higher building density generally leads to greater shadowing levels and larger Doppler spread, the urban layout affects multipath richness and channel dispersion in a complex  way, sometimes in contrast with common expectation and 3GPP standard models. Dense urban environments generally produce greater obstruction and richer multipath conditions at street level, whereas suburban layouts provide higher LoS probability but still allow scattering from more distant surfaces. As a result, the sensitivity of the Rician \(K\)-factor and delay spread to the specific urban morphology appears complex and case dependent.

Frequency dependence appears more limited, with similar propagation behavior within the S/C and Q/V bands, although slightly larger Rician \(K\)-factor values are observed at higher frequencies.

Finally, Doppler spread seems mostly driven by the mobility of the ground terminal compared to the motion of the satellite, which is actually much faster, but far less angularly dispersive at the same time. The achieved values of Doppler spread show that the channel remains time-invariant over intervals long enough to preserve the practical soundness of the average $K$-factor/delay spread characterization.

These findings contribute to the refinement of channel models for satellite components of future integrated terrestrial–non-terrestrial networks.


\section*{Appendix: Impact of a Finite System Bandwidth on the Delay Spread}

\setcounter{figure}{0}
\renewcommand{\thefigure}{A\arabic{figure}}

DS values reported in section III.C correspond to infinite-bandwidth power delay profiles, i.e., they are directly computed from each ray's simulated power and delay (as in \eqref{eq:DS}--\eqref{eq:DS_meandelay}), with no limitation on delay resolution other than the accuracy of the \gls{rt} simulator.

 Introducing a finite system bandwidth $B$, by contrast, makes the resulting DS system-dependent, as the transmitted signal imposes a finite delay resolution on the order of $1/B$, below which individual multipath components cannot be resolved.

To relate the DS values discussed throughout the paper to a specific finite-bandwidth system, we additionally compute a band-limited delay spread $\mathrm{DS}_B$ by filtering the channel response with an ideal low-pass filter of bandwidth $B=100$~MHz, representative of a typical wideband cellular channel, and applying the same DS definition of \eqref{eq:DS}--\eqref{eq:DS_meandelay} to the resulting continuous power delay profile.

\begin{figure}[t]
    \centering
    \includegraphics[width=0.9\linewidth]{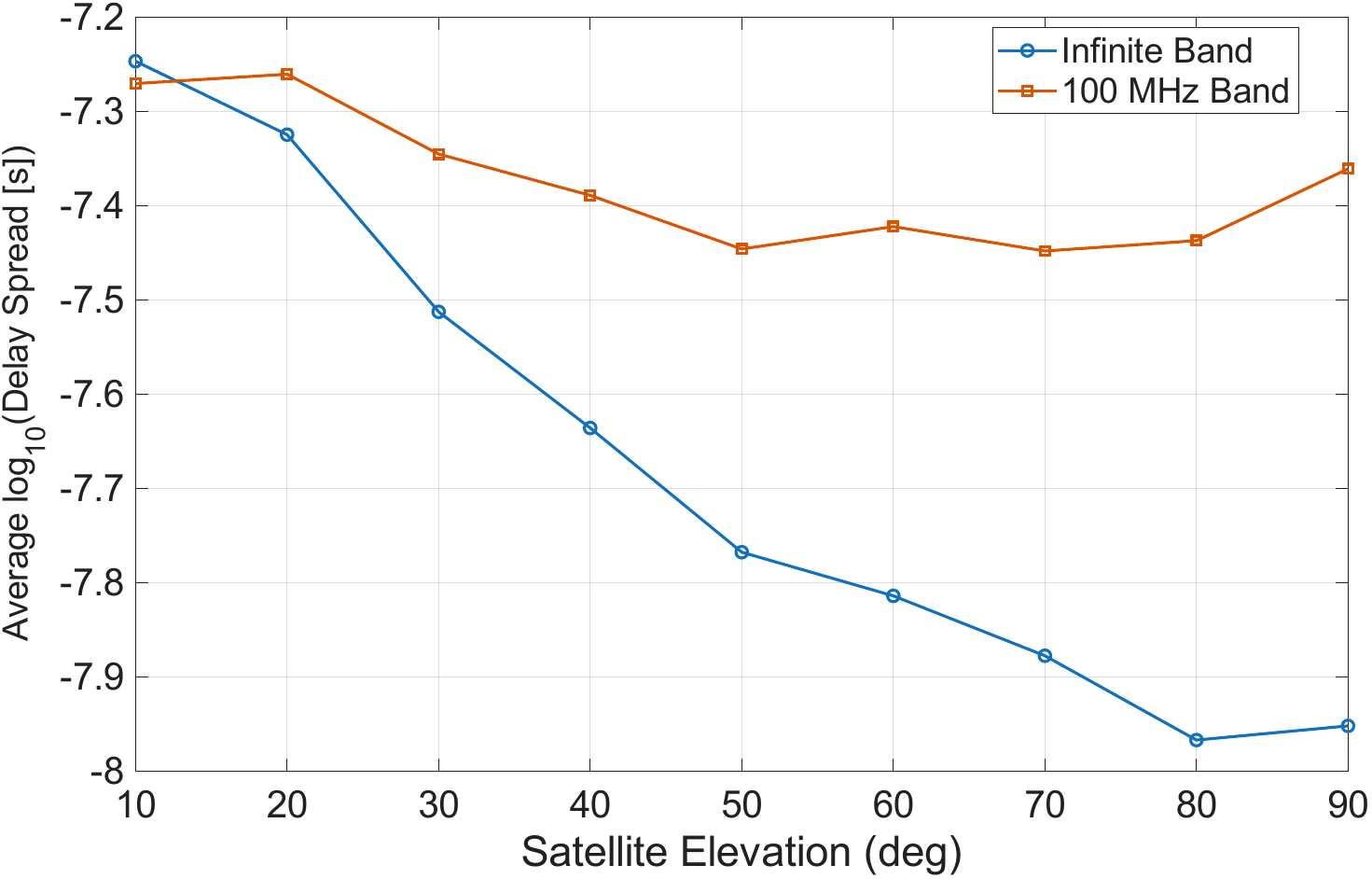}
    \caption{Average $\log_{10}(\mathrm{DS})$ vs. satellite elevation, infinite bandwidth vs. $B=100$~MHz, dense urban, 3.5~GHz, isotropic antenna.}
    \label{fig:DS_finiteband_3_5GHz_DU}
\end{figure}

\begin{figure}[t]
    \centering
    \includegraphics[width=0.9\linewidth]{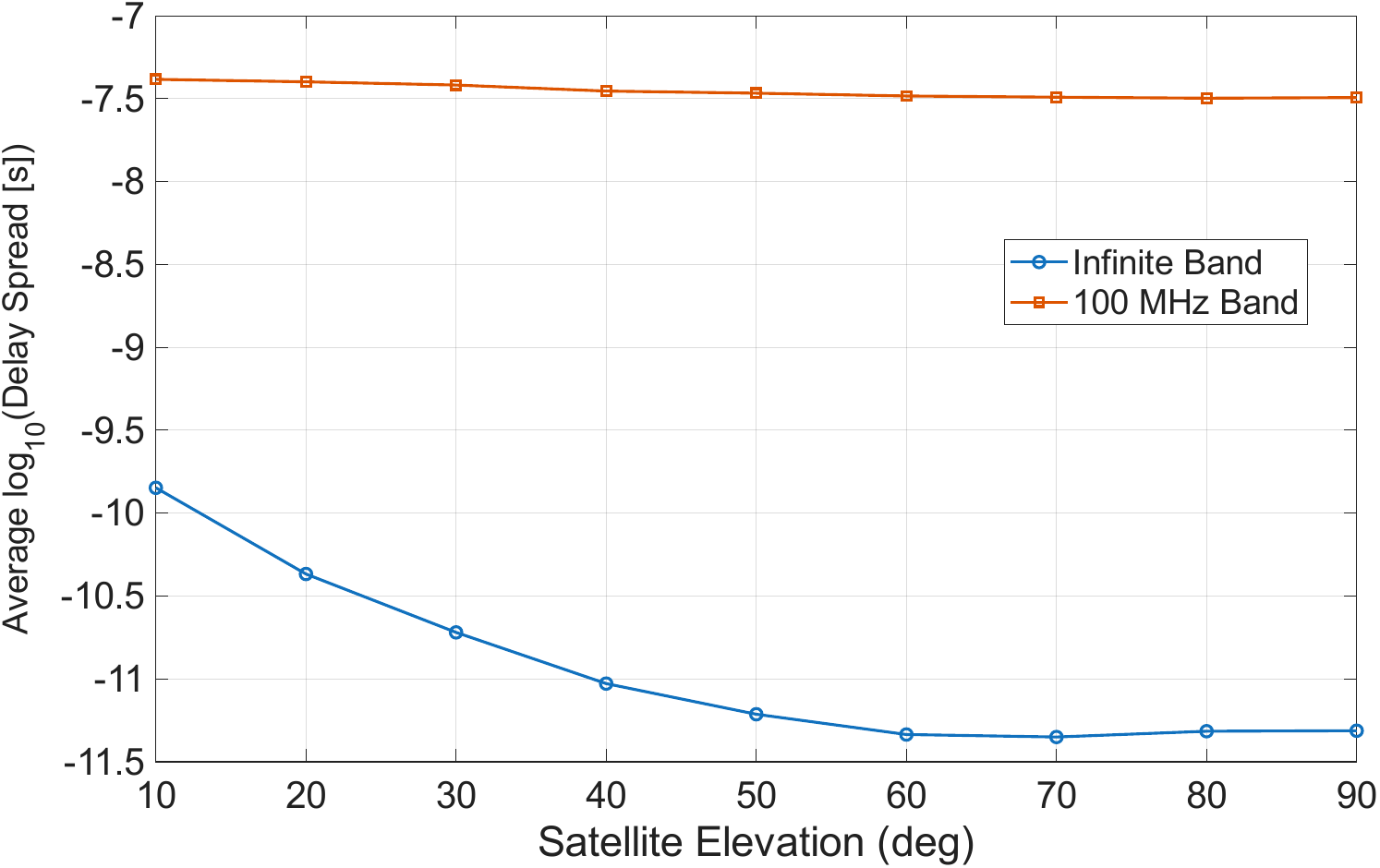}
    \caption{Average $\log_{10}(\mathrm{DS})$ vs. satellite elevation, infinite bandwidth vs. $B=100$~MHz, dense urban, 40~GHz, 60~cm directive antenna.}
    \label{fig:DS_finiteband_40GHz_DU}
\end{figure}

Figs.~\ref{fig:DS_finiteband_3_5GHz_DU} and \ref{fig:DS_finiteband_40GHz_DU} compare $\mathrm{DS}_B$ against the infinite-bandwidth DS in two cases of the dense urban scenario, both with the user equipment at street level: at 3.5~GHz with an isotropic antenna (Fig.~\ref{fig:DS_finiteband_3_5GHz_DU}), where the channel's intrinsic DS is larger, and at 40~GHz with the directive 60~cm antenna (Fig.~\ref{fig:DS_finiteband_40GHz_DU}), where intrinsic DS is much smaller.

 At 3.5~GHz, $\mathrm{DS}_B$ remains close to the intrinsic DS at low elevation, with a moderate difference emerging at higher elevation. At 40~GHz, by contrast, the intrinsic DS is much smaller, so that the resulting DS is essentially dominated by the limited system bandwidth.



\begin{figure}[t]
    \centering
    \includegraphics[width=0.9\linewidth]{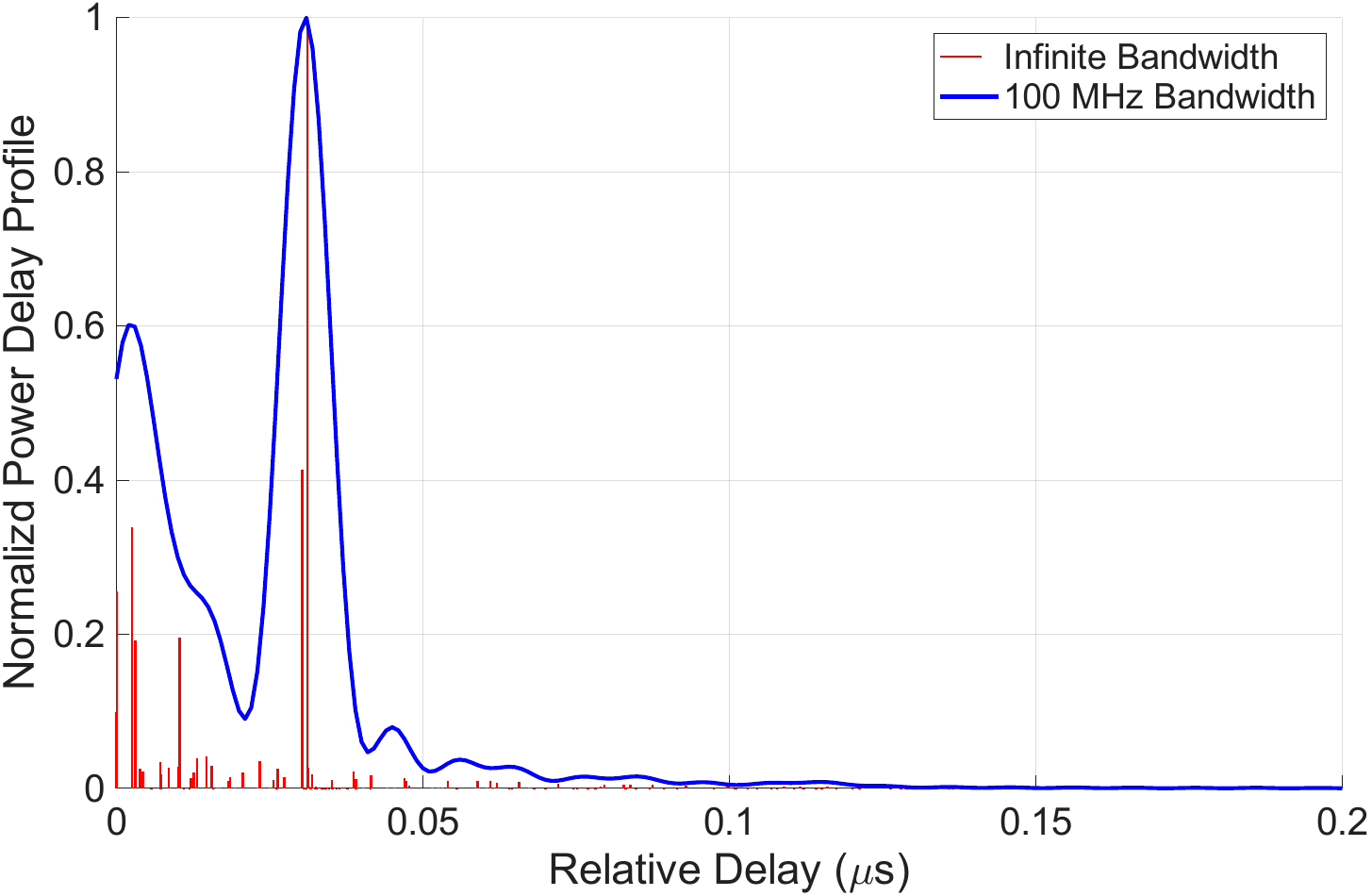}
    \caption{Example of power delay profile: infinite bandwidth (discrete spikes) vs. $B=100$~MHz band-limited (continuous). Obtained for the dense urban scenario, at 3.5 GHz and $20^\circ$ satellite elevation.}
    \label{fig:PDP_inf_vs_finite}
\end{figure}

Fig.~\ref{fig:PDP_inf_vs_finite} compares the infinite-bandwidth and band-limited power delay profiles for the dense urban scenario at 3.5~GHz and 20$^\circ$ elevation in an NLoS case. The earliest rays are rooftop diffractions, followed by diffuse-scattering contributions, while the dominant peak is due to a specular reflection from a more distant wall.

Overall, this confirms that the infinite-bandwidth DS provides the relevant, system-independent characterization of the propagation channel. A finite system bandwidth introduces an additional system-dependent contribution that can dominate the observed DS when the intrinsic DS is very small, as in LoS, rooftop-level, high-frequency links.

\bibliographystyle{IEEEtran}
\bibliography{biblio.bib}

\end{document}